\documentclass[journal,twoside]{IEEEtranTCOM}

\IEEEoverridecommandlockouts

\usepackage{cite}

\usepackage{graphicx}
\usepackage{subfigure}
\usepackage{wrapfig}
\usepackage{comment}

\usepackage{epstopdf}
\usepackage[cmex10]{amsmath}
\usepackage{amsmath}
\usepackage{amsthm}
\usepackage{amsfonts}	

\usepackage{algorithmic}
\usepackage[ruled,vlined,boxed,linesnumbered]{algorithm2e}

\usepackage{url}

\usepackage[all]{xy}

\def \single {single}

\def \columnmode {double}

\def \thesis {thesis}
\def \journal {journal}

\def \thesismode {journal}

\ifx \thesismode \thesis

	\newcommand\secref{Ch.~\ref}
	\newcommand\appref{Ch.~\ref}

	\newcommand\convbsc{Thm.~\ref{thm:converse_bsc}}
	\newcommand\convawgn{Thm.~\ref{thm:converse_awgn}}

	\newcommand\paperref{chapter}

\else

	\newcommand\secref{Sec.~\ref}
	\newcommand\appref{Appendix~\ref}

	\newcommand\convbsc{\cite[Theorem 6]{Polyanskiy_IT_2011_NonAsym}}
	\newcommand\convawgn{\cite[Theorem 4]{Polyanskiy_IT_2011_NonAsym}}

	\newcommand\paperref{paper}

\fi

\renewcommand{\P}{\mathrm{P}}

\newcommand{\E}{\mathrm{E}}

\newtheorem{theorem}{Theorem}
\newtheorem{definition}{Definition}

\newtheorem{corollary}{Corollary}

\usepackage{times,color}

\begin{document}
%
\title{Variable-length Convolutional Coding for \\ Short Blocklengths with Decision Feedback}	


\author{Adam~R.~Williamson,~\IEEEmembership{Member,~IEEE,}
        Tsung-Yi~Chen,~\IEEEmembership{Member,~IEEE,}
        and~Richard~D.~Wesel,~\IEEEmembership{Senior~Member,~IEEE}
\thanks{
\ifx \columnmode \single
   A. R. Williamson is with Northrop Grumman, Redondo Beach, CA 90278 USA (adam.williamson.us@ieee.org). R. D. Wesel is with the EE Dept., UCLA, Los Angeles, CA 90095 USA (wesel@ee.ucla.edu). T.-Y. Chen is with the Dept. of EECS, Northwestern University, Evanston, IL 60208 USA (tsungyi.chen@northwestern.edu). A. R. Williamson and T.-Y. Chen were with UCLA when this research was conducted.
\else
   A. R. Williamson is with Northrop Grumman, Redondo Beach, CA 90278 USA (adam.williamson.us@ieee.org). R. D. Wesel is with the Electrical Engineering Department, University of California, Los Angeles, CA 90095 USA (wesel@ee.ucla.edu). T.-Y. Chen is with the Department of Electrical Engineering and Computer Science, Northwestern University, Evanston, IL 60208 USA (tsungyi.chen@northwestern.edu).  A. R. Williamson and T.-Y. Chen were with UCLA when this research was conducted.
\fi
\ifx \columnmode \single
   Portions of this work were presented at the 2013 Int. Symp. on Inf. Theory (ISIT) in Istanbul, Turkey.
\else
   Portions of this work were presented at the 2013 International Symposium on Information Theory (ISIT) in Istanbul, Turkey.
\fi
This material is based upon work supported by the National Science Foundation under Grant Number 1162501. Any opinions, findings, and conclusions or recommendations expressed in this material are those of the author(s) and do not necessarily reflect the views of the National Science Foundation.
}
}

\maketitle

\ifx \thesismode \journal
\ifx \columnmode \single
   \vspace{-24pt}
   \vspace{-24pt}
\fi
\fi

%
%
\begin{abstract}

This \paperref~presents a variable-length decision-feedback coding scheme that achieves  high rates at short blocklengths. This scheme uses the Reliability-Output Viterbi Algorithm (ROVA) to determine when the receiver's decoding estimate satisfies a given error constraint. We evaluate the performance of both terminated and tail-biting convolutional codes at average blocklengths less than 300 symbols, using the ROVA and tail-biting ROVA, respectively.
Comparing with recent results from finite-blocklength information theory, simulations for both the BSC and the AWGN channel show that the reliability-based decision-feedback scheme can surpass the random-coding lower bound on throughput for feedback codes at some blocklengths less than 100 symbols.  This is true both when decoding after every symbol is permitted, and when decoding is limited to a small number of increments.  
Finally, the performance of the reliability-based stopping rule with the ROVA is compared to retransmission decisions based on CRCs. For short blocklengths where the latency overhead of the CRC bits is severe, the ROVA-based approach delivers superior rates.

\end{abstract}

\ifx \columnmode \single
\else
\fi

\IEEEpeerreviewmaketitle

%
%
\section{Introduction}
\label{sec:vlf_intro}

%
\subsection{Overview and Related Literature}
\label{sec:relatedliterature}
%


Despite Shannon's 1956 result \cite{Shannon_IT_1956} that noiseless feedback does not increase the asymptotic capacity of point-to-point, memoryless channels,
feedback has other benefits for these channels that have made it a staple in modern communication systems. For example, feedback can simplify encoding and decoding operations and has been incorporated into incremental redundancy (IR) schemes proposed at least as early as 1974 \cite{Mandelbaum_IT_1974}. Hagenauer's introduction of rate-compatible punctured convolutional (RCPC) codes allows the same encoder to be used under varying channel conditions with feedback determining when to send additional coded bits \cite{Hagenauer_TCOM_1988}.  Additionally, feedback can significantly improve the error exponent, which governs the error probability as a function of blocklength (see, e.g., \cite{Burnashev_1976}). 

Perhaps the most important benefit of feedback is its ability to reduce the average blocklength required to approach capacity.
The pioneering work of Strassen \cite{Strassen_1962} analyzed the backoff from capacity at finite blocklengths without feedback.
More recently, a number of papers have extended these results, most notably Hayashi \cite{Hayashi_Info_Spectrum_TransIT_2009} and Polyanskiy, Poor and Verd\'{u} \cite{Polyanskiy_CCR_2010}. Polyanskiy et al. \cite{Polyanskiy_CCR_2010} demonstrated that short blocklengths impose a severe penalty on the maximum achievable rate, showing that even when the best fixed-length block code is paired with an Automatic Repeat reQuest (ARQ) strategy, the maximum rate is slow to converge to the asymptotic (Shannon) capacity.
However, when variable-length coding is used on channels with noiseless feedback, the maximum rate improves dramatically for short (average) blocklengths \cite{Polyanskiy_IT_2011_NonAsym}.

In \cite{Polyanskiy_IT_2011_NonAsym}, Polyanskiy et al. formalize the notion of both variable-length feedback (VLF) codes and variable-length feedback codes {\em with termination} (VLFT). In VLF schemes the receiver decides when to stop the transmission and informs the transmitter of that decision using feedback.  In VLFT  schemes, the transmitter decides when to stop the transmission based on information it receives from the receiver via feedback, such as full information about the received symbols or what tentative decoding decision the receiver has made.  With VLFT, the transmitter informs the receiver of the decision to stop transmitting using a special noiseless termination symbol on a separate channel.  The VLFT construct of a noiseless termination symbol in [8] essentially creates a genie-aided decoder that can stop as soon as its tentative decoding decision is correct.  This obviates the need for either Cyclic Redundancy Checks (CRCs) or reliability-based retransmission decisions.

%
%
In contrast, this paper investigates how realistic error detection impacts the achievable rates at short blocklengths.  Thus the focus is on VLF codes.  Our schemes employ {\em decision} feedback, for which feedback is only used to inform the transmitter when to stop. This is in contrast to {\em information}-feedback VLF codes, which allow the transmitter to adapt its transmission based on information about the previously received symbols. See, e.g., \cite{Williamson_2phase_ITA_2013,Vakilinia_LDPC_feedback_ISIT_2014,Vakilinia_LDPC_feedback_ITW_2014}, for recent investigations of information feedback at short blocklengths. 
Our work is the first to specifically address the question of how closely existing coding techniques using feedback (as opposed to theoretical constructs such as random coding) can approach capacity as a function of the average blocklength, without  a special noiseless transmission or a genie informing the receiver when it has correctly decoded. We look in the region of short average blocklength where VLF lower bounds  in \cite{Polyanskiy_IT_2011_NonAsym} are well below the channel capacity.  Moreover, our work is the first to focus design and optimization efforts towards achieving the best possible VLF performance in this region.

Previously, Chen et al. \cite{Chen_ISIT_2013,Chen_Feedback_Journal_2013} studied practical implications of VLFT codes such as the effects of periodic decoding (i.e., only decoding and sending feedback after every $I > 1$ symbols). The
analysis in \cite{Chen_ISIT_2013} and \cite{Chen_Feedback_Journal_2013} used IR based on finite-length mother codes, showing that a length-$N$ mother code can be used in a variable-length transmission scheme and provide the same performance as an infinite-length code (as in \cite{Polyanskiy_IT_2011_NonAsym}), up to second order terms. The importance of these results is that ``good" finite-length codes can still achieve rates approaching capacity at short blocklengths.
Contemporaneous with \cite{Polyanskiy_CCR_2010} and \cite{Polyanskiy_IT_2011_NonAsym}, Chen et al. \cite{Chen_2011_ICC, Chen_2011_ITA} showed that the relatively simple decoding of short-blocklength convolutional codes in an IR setting could match the throughput delivered by long-blocklength turbo codes.


%
%
Prior to the definition of VLF and VLFT codes, IR and hybrid ARQ (HARQ) had been discussed extensively in the communication literature (e.g., \cite{Costello_HARQ_IT_1998,Visotsky_RBIR_TCOM_2005, Lott_Soljanin_ITW_2007,Fricke_Reliability_HARQ_TCOM_2009}).
Costello et al. \cite{Costello_HARQ_IT_1998} provides an overview of HARQ (i.e., the combination of error correction and ARQ) and discusses various applications.
Lott et al. \cite{Lott_Soljanin_ITW_2007} provides a survey of HARQ work until 2007.
A partial description of these and more recent HARQ schemes follows. We refer to schemes that rely on a genie-aided decoder as VLFT schemes, whereas schemes that implement a receiver-based retransmission rule are categorized as VLF schemes.
For example, using a CRC at the receiver qualifies as a VLF scheme.


The reliability-based HARQ scheme in Shea \cite{Shea_HARQ_EL_2002} uses VLFT codes based on a concatenation of turbo and block parity-check codes. The transmitter uses information feedback of the symbol posterior probabilities to determine which coded symbols to send in subsequent transmissions.

Convolutional codes (CCs) are commonly used in HARQ schemes. Roongta and Shea \cite{Roongta_HARQ_ICC_2003,Roongta_HARQ_WCNC_2004} present a reliability-based HARQ scheme using VLFT codes based on CCs. Information feedback of the symbol posterior probabilities, obtained from the BCJR algorithm \cite{BCJR_TransIT_1974}, determines subsequent transmissions, as in \cite{Shea_HARQ_EL_2002}.
Raghavan and Baum \cite{Raghavan_ROVA_TransIT_1998} and Fricke and Hoeher \cite{Fricke_HARQ_NAW_2006,Fricke_Approx_ROVA_VTC_2007,Fricke_Reliability_HARQ_TCOM_2009} present reliability-based type-I HARQ schemes using terminated CCs in a VLF decision-feedback setting. In \cite{Fricke_Reliability_HARQ_TCOM_2009}, the reliability-based scheme is also compared to a code-based scheme using CRCs for error detection.
Visotsky et al. \cite{Visotsky_RBIR_TCOM_2005} introduce a type-II HARQ scheme (i.e., IR) using CCs and a reliability-based retransmission rule. In this case, the reliability metric is based on the average magnitude of the log-likelihood ratios of the source symbols. The transmission strategy maximizes throughput subject to a delay constraint.

Pai et al. \cite{Pai_HARQ_TBCC_TransVT_2011} uses tail-biting convolutional codes (TBCCs) for type-I HARQ in a VLF decision-feedback setting. In Pai et al. \cite{Pai_HARQ_TBCC_TransVT_2011}, TBCCs are decoded with a sub-optimal decoder, whereas in this paper we use an optimal (ML) decoder for TBCCs.


A number of reliability-based decoding algorithms have been developed that take advantage of the trellis structure of CCs. The Yamamoto-Itoh algorithm \cite{Yamamoto_Itoh_Algo_TransIT_1980} for terminated CCs computes a reliability measure for the decoded sequence by comparing the metric differences between merging branches in the trellis. 
In \cite{Pai_HARQ_TBCC_TransVT_2011}, Pai et al. extend the Yamamoto-Itoh algorithm to handle TBCCs. 
In both \cite{Yamamoto_Itoh_Algo_TransIT_1980} and \cite{Pai_HARQ_TBCC_TransVT_2011}, the reliability  measure is different from the word-error probability, however, and is not sufficient to guarantee a particular undetected-error probability.

In contrast, Raghavan and Baum's Reliability-Output Viterbi Algorithm (ROVA) for terminated CCs \cite{Raghavan_ROVA_TransIT_1998} and Williamson et al.'s TB ROVA for TBCCs \cite{Williamson_TBROVA_2014} 
compute the posterior probability of the codeword exactly. Fricke and Hoeher \cite{Fricke_Approx_ROVA_VTC_2007} present an approximate method to compute the posterior probability for terminated CCs.

Hof et al. \cite{Hof_Conv_Bounds_ISCTA_2009} modify the Viterbi algorithm to permit generalized decoding according to Forney's generalized decoding rule \cite{Forney_Erasure_TransIT_1968}.
When the generalized decoding threshold is chosen for maximum likelihood (ML) decoding with erasures and the erasure threshold is chosen appropriately, this augmented Viterbi decoder is equivalent to the ROVA.

The ROVA and its extensions are different from the well-known BCJR algorithm \cite{BCJR_TransIT_1974} and its tail-biting counterparts \cite{Anderson_TB_MAP_JSAC_1998} and \cite[Ch. 7]{Johannesson_Fundamentals_Conv_1999}, which compute the posterior probabilities of individual source symbols. Instead, the ROVA computes the posterior probability of the entire decoded sequence.

Soljanin et al. \cite{Soljanin_LDPC_HARQ_2005,Soljanin_LDPC_HARQ_ITW_2006} present an HARQ scheme based on low-density parity-check (LDPC) codes with random transmission assignments and use an ML decoding analysis to determine how many incremental symbols to send after each failed transmission.
The performance of the LDPC-based scheme is compared to a second decision-feedback VLFT coding scheme using Raptor codes.
Soljanin et al. also study LDPC-based HARQ over a time-varying binary erasure channel in \cite{Soljanin_LDPC_ITW_2009,Soljanin_LDPC_IT_2012}.

Pfletschinger et al. \cite{Pfletschinger_LDPC_HARQ_TWC_2014} use rate-adaptive, non-binary LDPC codes in a type-II HARQ scheme over the Rayleigh fading channel in the VLFT setting. In \cite{Pfletschinger_LDPC_HARQ_TWC_2014} Pfletschinger et al.  present two blocklength-optimization algorithms that seek to maximize the throughput, subject to an overall error constraint. One scenario uses decision feedback and selects blocklengths based on channel statistics and the other uses information feedback to adaptively select blocklengths based on accumulated mutual information. The information-feedback approach is referred to as ``multibit NACK" or ``intelligent NACK".

Rateless spinal codes are promising candidates for high-throughput HARQ protocols. In \cite{Perry_Spinal_Hotnets_2011,Perry_Spinal_Sigcomm_2012}, Perry et al. present spinal codes, which are nonlinear and use pseudo-random hash functions to produce a rateless sequence of coded symbols. Simulations in a decision-feedback VLFT setting  show that spinal codes outperform Raptor and Strider codes. In \cite{Romero_MS_Thesis_2014}, Romero evaluates the performance of spinal codes in a decision-feedback VLF setting by adding CRCs for error detection.

Chen et al. \cite{Chen_Polar_HARQ_TCOM_2013} show that rate-compatible polar codes in a VLFT setting over the BI-AWGN channel can perform as well as existing HARQ schemes based on turbo codes and LDPC codes. In \cite{Chen_Polar_HARQ_WCNC_2014}, Chen et al. introduce an HARQ scheme over the Rayleigh fading channel that uses Chase combining of polar codes.

In this paper, we focus on latency under $300$ bits and evaluate information blocklengths as low as $k$$=$$8$ bits, whereas the existing HARQ work does not focus on such short-blocklength performance. 
%
Only the following papers evaluate information blocklengths $k$ under 300 bits: \cite{Fricke_Reliability_HARQ_TCOM_2009} ($k$$=$$40$ bits), \cite{Mukhtar_Turbo_Product_TCOM_2014} ($k$$=$$121$ bits), \cite{Soljanin_LDPC_HARQ_2005} ($k$$=$$154$ bits), \cite{Fricke_HARQ_NAW_2006} ($k$$=$$192$ bits), and \cite{Pai_HARQ_TBCC_TransVT_2011} ($k$$=$$288$ bits).
Due to the type-I HARQ structure with rate-$1/2$ CCs in \cite{Fricke_Reliability_HARQ_TCOM_2009}, \cite{Fricke_HARQ_NAW_2006} and \cite{Pai_HARQ_TBCC_TransVT_2011}, the maximum throughput possible is only $0.5$ bits per channel use and the minimum latency is $2k$ bits.
%
Mukhtar et al. \cite{Mukhtar_Turbo_Product_TCOM_2014} introduces a CRC-free HARQ scheme using turbo product codes, showing how using extended BCH codes as component codes can provide inherent word-error detection and avoid the rate loss of CRCs. Throughput results in \cite{Mukhtar_Turbo_Product_TCOM_2014} are not given for the $k$$=$$121$ case, however, so the smallest $k$ with throughput results is $k$$=$$676$ bits.
For some SNRs, the LDPC-based HARQ scheme in Soljanin et al. \cite{Soljanin_LDPC_HARQ_2005} provides latency under $300$ bits, but with lower throughput than our CC-based scheme.
We will show in \secref{sec:stop_feedback} that our scheme can achieve throughputs of approximately $0.53$ bits per channel use for the 2 dB Gaussian channel with a latency of $121$ bits, whereas LDPC codes in \cite{Soljanin_LDPC_HARQ_2005} provide similar rates with a latency of approximately 300 bits and with substantially higher word-error probability.

%
%


Makki et al. \cite{Makki_Finite_HARQ_WLC_2014} analyzes the achievable throughput of type-II HARQ systems by extending finite-blocklength results for the Rayleigh fading channel from \cite{Polyanskiy_CCR_2010,Yang_fading_ISIT_2013,Yang_fading_TransIT_2014} to rate-compatible code families. Numerical examples using either $k$$=$$300$ or $k$$=$$600$ information symbols that HARQ can improve throughput by up to 15\% versus fixed-length communication without feedback.
Similarly, \cite{Makki_Green_ARQ_Globecom_2014} uses finite-blocklength results from \cite{Polyanskiy_CCR_2010,Yang_fading_ISIT_2013,Yang_fading_TransIT_2014} to study the effects of power allocation for type-I ARQ systems with Rayleigh fading.
As in \cite{Polyanskiy_CCR_2010}, the information-theoretic analysis in both \cite{Makki_Finite_HARQ_WLC_2014} and \cite{Makki_Green_ARQ_Globecom_2014} does not consider explicit code constructions like those in this paper.
%

Several recent papers \cite{Hehn_LDPC_v_Convolutional_TCOM_2009,Maiya_Costello_Low_Lat_Coding_2012,Rachinger_Low_Lat_Coding_2014}  compare the short-blocklength performance of fixed-length convolutional and LDPC codes (i.e., without feedback). In \cite{Rachinger_Low_Lat_Coding_2014}, for example, the authors investigate the SNR required to reach a given BER or FER for each code, as a function of the blocklength. Rachinger et al. \cite{Rachinger_Low_Lat_Coding_2014} includes finite-blocklength bounds from \cite{Polyanskiy_CCR_2010}. These papers show that convolutional codes perform best for short blocklengths.

\subsection{Contributions}
\label{sec:contribs}

This paper seeks to maximize the throughput of variable-length feedback (VLF) codes  with average blocklengths less than 300 symbols while meeting an error probability constraint.  Our schemes use CCs (due to their excellent performance at short blocklengths) in  explicit decision-feedback coding schemes that surpass the random-coding lower bound in \cite{Polyanskiy_IT_2011_NonAsym}.  
Simulation results demonstrate this performance for the binary symmetric channel (BSC) and binary-input additive white Gaussian noise (BI-AWGN) channel.


As in our precursor conference paper \cite{Williamson_ROVA_ISIT_2013}, our first decision-feedback scheme uses Raghavan and Baum's ROVA \cite{Raghavan_ROVA_TransIT_1998} for terminated CCs to compute the posterior probability of the decoded word and stops transmission when that word is sufficiently likely.
While this scheme delivers high rates at moderate blocklengths, the termination bits introduce non-negligible rate loss at the shortest blocklengths. 
%
In exchange for increased decoding complexity compared to terminated convolutional codes, TB codes do not suffer from rate loss at short blocklengths. 
%
Our second decision-feedback scheme uses the TB ROVA to avoid the overhead of termination bits.  Both the ROVA and the TB ROVA allow the decoder to request retransmissions without requiring parity bits to be sent for error detection.  For completeness we compare these schemes with an approach using CRCs. 

When delay constraints or other practical considerations preclude decoding after every symbol, decoding after groups of  symbols (packets) is required, and the incremental transmission lengths must be optimized.  \appref{sec:blocklength_algo} provides a numerical optimization algorithm for selecting the $m$ optimal blocklengths in a general $m$-transmission IR scheme. \appref{sec:algo_tb_rova} particularizes this algorithm to the reliability-based scheme using the TB ROVA.


The new contributions relative to our precursor conference paper \cite{Williamson_ROVA_ISIT_2013} are as follows: this \paperref~incorporates the TB ROVA of \cite{Williamson_TBROVA_2014}, investigates the performance of ``packet'' transmissions by introducing a novel blocklength-selection algorithm, compares ROVA-based retransmission to CRC-based retransmission, and extends bounds in \cite{Polyanskiy_IT_2011_NonAsym} to repeat-after-$N$ codes.
The remainder of this \paperref~proceeds as follows: \secref{sec:vlf_notation} introduces relevant notation. 
\secref{sec:vlf} reviews the fundamental limits for VLF codes from \cite{Polyanskiy_IT_2011_NonAsym} and presents extensions of the random-coding lower bound to VLF systems with ``packets''.
\secref{sec:stop_feedback} evaluates the performance of ROVA-based and CRC-based stopping rules in the decision-feedback setting.
\secref{sec:vlf_conc} concludes the \paperref.

%
\subsection{Notation}
\label{sec:vlf_notation}

In general, capital letters denote random variables and lowercase letters denote their realizations (e.g., random variable $Y$ and value $y$).
Superscripts denote vectors unless otherwise noted, as in $y^\ell = (y_1, y_2, \dots, y_\ell)$, while subscripts denote a particular element of a vector: $y_i$ is the $i$th element of $y^\ell$. 
The expressions involving $\log(~)$ and $\exp\{~\}$ in information-theoretic derivations are valid for any base, but numerical examples use base 2 and present results in units of bits.

%
%
\section{VLF Coding Framework}
\label{sec:vlf_background}
\label{sec:vlf}

%
\subsection{Definitions for VLF Codes}


%


For finite-length block codes without feedback, Polyanskiy et al. \cite{Polyanskiy_CCR_2010} provide achievability (lower) and converse (upper) bounds on the maximum rate, along with a normal approximation of the information density that can approximate both bounds for moderate blocklengths. 
In contrast to this tight characterization of the no-feedback case, there is a large gap between the lower and upper bounds for VLF codes at short average blocklengths presented in Polyanskiy et al. \cite{Polyanskiy_IT_2011_NonAsym}.   
This section reviews the fundamental limits for VLF codes from \cite{Polyanskiy_IT_2011_NonAsym} and provides extensions of the lower bound to repeat-after-$N$ codes, which are similar in principle to the finite-length VLFT codes studied in \cite{Chen_ISIT_2013,Chen_Feedback_Journal_2013}. This framework will allow us to evaluate the short-blocklength performance of the decision-feedback schemes in \secref{sec:stop_feedback} in terms of these fundamental limits.

We assume there is a noiseless feedback channel.
The noisy forward channel is memoryless, has input alphabet ${\mathcal X}$ and has output alphabet ${\mathcal Y}$. The channel satisfies
\begin{align}
	\P_{Y_n | X^n,Y^{n-1}} & (a | b^n, c) = \P_{Y_n|X_n} (a | b_n) = \P_{Y_1|X_1} (a | b_1) \\
	&~\forall~a \in {\mathcal Y}, b^n \in {\mathcal X}^n, c \in {\mathcal Y}^{n-1}, \text{ and } \forall ~n \in {\mathbb Z}^{+} . \nonumber
\end{align}
A discrete, memoryless channel (DMC) is a special case when $\mathcal X$ and $\mathcal Y$ are countable.

\begin{definition}	\label{def:vlf}
(From \cite{Polyanskiy_IT_2011_NonAsym}) An ($\ell, M, \epsilon)$ \textbf{variable-length feedback (VLF)} code is defined by:
\begin{itemize}
	\item A message $W \in {\mathcal W} = \{1, \dots, M\}$, assumed to be equiprobable. (The positive integer $M$ is the cardinality of the message set $\mathcal W$.)
%
	\item A random variable $U \in {\mathcal U}$ and a probability distribution $P_U$ on the space ${\mathcal U}$.
$U$ represents common randomness that is revealed to both transmitter and receiver before communication begins, which facilitates the use of random-coding arguments in the sequel.
%
	\item A sequence of encoder functions $f_n : {\mathcal U} \times {\mathcal W} \times {\mathcal Y}^{n-1} \rightarrow 
{\mathcal X}, n \geq 1$, which defines the $n$th channel input:
\begin{align}
	X_n = f_n(U, W, Y^{n-1}).
\end{align}
%
	\item A sequence of decoder functions $g_n : {\mathcal U} \times {\mathcal Y}^n \rightarrow {\mathcal W}, n \geq 1$, providing an estimate $\hat W_n$ of the message $W$:
%
\begin{align}
	\hat W_n =  g_n(U, Y^n).\label{eq:Wnestimates}
\end{align}
	\item An integer-valued random variable $\mu \geq 0$, which is a stopping time of the filtration ${\mathcal G}_n~=~\sigma \{U, Y_1, \dots, Y_n\}$. The stopping time satisfies
\begin{align}
	\E[\mu] \leq \ell.
	\label{eqn:tau_ell}
\end{align}
	\item A final decision computed at time $\mu$, at which the error probability must be less than $\epsilon$ $(0 \leq \epsilon \leq 1)$: 
\begin{align}
	\P[\hat W_\mu \neq W] \leq \epsilon.
\end{align}
\end{itemize}
\end{definition}
%

\begin{figure}
\ifx \columnmode \single
   \centering \def\svgwidth{400pt}
   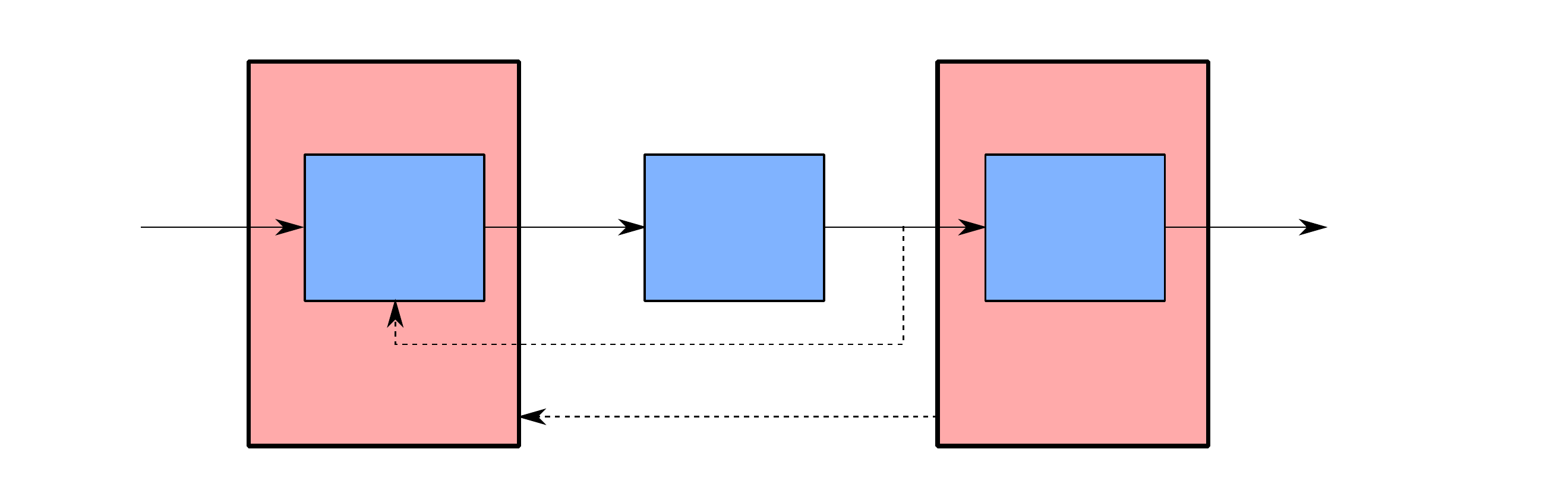
\else
   \centering \def\svgwidth{270pt}
   \input{willi1.pdf_tex}
\fi
\caption{Illustration of the VLF coding framework, with message $W$, encoder (Enc.) output $X_n$ at time $n$, memoryless channel output $Y_n$, and decoder (Dec.) estimate $\hat W_n$. The feedback link (dashed line) to the encoder is assumed to be noiseless. With decision feedback, the transmitter disregards the received symbols $Y_n$ and only uses feedback to determine whether to stop (i.e., when $\mu = n$).}
\ifx \columnmode \single
   \vspace{-30pt}
\fi
\label{fig:vlf_block}
\end{figure}

In the definition above, the receiver attempts to decode after each received symbol. Because the final decision is not computed until time $n = \mu$, the estimates $\hat W_n$ in \eqref{eq:Wnestimates} for $n < \mu$ may be considered tentative estimates that do not affect the decoding outcome. This differs slightly from \cite{Polyanskiy_IT_2011_NonAsym}, which does not require the decoder to compute the tentative estimates $g_n(U, Y^n)$ for $n<\mu$.
We include the definition of $\hat W_n$ here for consistency with the reliability-based stopping approach in \secref{sec:stop_feedback}.
The receiver uses some stopping rule (to be specified in \secref{sec:stop_feedback}) to determine when to stop decoding and informs the transmitter via feedback.
This VLF coding framework is illustrated in Fig.~\ref{fig:vlf_block}.
Eq. \eqref{eqn:tau_ell} indicates that for an $(\ell, M, \epsilon)$ VLF code, the expected length will be no more than $\ell$. The average rate $R$ is given as $R = \frac{\log M}{E[\mu]}$ and is lower-bounded by $\frac{\log M}{\ell}$.

Polyanskiy et al. \cite{Polyanskiy_IT_2011_NonAsym} define a class of VLF codes called {\em stop-feedback codes} that satisfy:
\begin{align}
	f_n(U, W, Y^{n-1}) = f_n(U, W).
	\label{eqn:stop-feedback}
\end{align}
For stop-feedback codes, the encoded symbols are independent of the previously received noisy-channel outputs. The feedback link is used only to inform the transmitter when the receiver has determined that the transmission should terminate. This paradigm was referred to as {\em decision feedback} in early papers by Chang \cite{Chang_information_feedback_1956} and by Forney \cite{Forney_Erasure_TransIT_1968}, the term which we will use in the sequel, and as a channel used without feedback by Massey \cite{Massey_Causality_ISITA_1990}.

Decision-feedback codes are practically relevant because they require at most one bit of feedback per forward channel use (per decoding attempt) for the receiver to communicate its termination decision (i.e., to stop or to request additional symbols).
This feedback bit is usually referred to as an ACK (acknowledgment) or NACK (negative acknowledgment).  The entropy of this feedback bit is typically much less than one bit per symbol since only a single ACK is transmitted per message.  In practical implementations the feedback channel may remain silent until sending the ACK signal.  Also, for the practically important case of grouped messages, a single bit is transmitted per group (per decoding attempt).  In \secref{sec:stop_feedback} we will demonstrate examples in which the number of groups transmitted is rarely above five.
A consequence of the decision-feedback restriction is that the codeword corresponding to message $W$, $\{X_n(W)\}_{n=1}^\infty$, can be generated before any symbols are transmitted over the channel, as the codeword does not depend on the realizations of $\{Y_n\}_{n=1}^\infty$.

In contrast, VLF codes for which \eqref{eqn:stop-feedback} is not generally true are referred to as information-feedback codes.  The most general form of information feedback is for the receiver to send each of its received symbols $Y_n$ to the transmitter.  With information feedback, the transmitter may adapt its transmissions based on this feedback, directing the receiver to the correct codeword.
Naghshvar et al. \cite{Naghshvar_binary_inputs_ITW_2012,Naghshvar_EJS_arxiv_2013} discuss this type of feedback in the context of active sequential hypothesis testing and demonstrate a deterministic, sequential coding scheme for DMCs that achieves the optimal error-exponent.

%
%
\subsection{Fundamental Limits for VLF Codes}

The VLF converse bounds referenced in this \paperref~come from Polyanskiy et al. \cite[Theorem 4]{Polyanskiy_IT_2011_NonAsym}
and \cite[Theorem 6]{Polyanskiy_IT_2011_NonAsym}. The latter \cite[Theorem 6]{Polyanskiy_IT_2011_NonAsym} applies only to channels with finite maximal relative entropy $C_1$, but provides a tighter upper bound than \cite[Theorem 4]{Polyanskiy_IT_2011_NonAsym}.

The following VLF achievability theorems make use of the information density at blocklength $n$, $i(x^n; y^n)$, defined as 
%
\begin{align}
	i(x^n; y^n) = \log \frac{d\P(Y^n = y^n|X^n = x^n)}{d\P(Y^n = y^n)}.
\end{align}
	\begin{theorem}[Random-coding lower bound{\cite[Theorem 3]{Polyanskiy_IT_2011_NonAsym}}]
	For a scalar $\gamma > 0$, $\exists$ an $(\ell, M, \epsilon)$ VLF code satisfying
	\begin{IEEEeqnarray}{rCl}
	\ell &\leq& \mathrm{E}[\tau], 	\label{eqn:vlf_Etau} \\	
	\epsilon &\leq& (M-1) \mathrm{P}[ \bar\tau \leq \tau],	\label{eqn:vlf_epsilon}
	\end{IEEEeqnarray}
	where $\gamma$ is a threshold for the hitting times $\tau$ and $\bar\tau$:
	\begin{IEEEeqnarray}{rCl}
	\tau &=& \inf \{n \geq 0 : i(X^n;Y^n) \geq \gamma \} 	\label{eqn:vlf_tau} \\
	\bar\tau &=& \inf \{n \geq 0 : i(\bar X^n;Y^n) \geq \gamma \} 	\label{eqn:vlf_tau_bar},
	\end{IEEEeqnarray}
	and where $\bar X^n$ is distributed identically to $X^n$, but is independent of $(X^n, Y^n)$.
	\label{thm:achievability}
	\end{theorem}
\ifx \columnmode \single
   \vspace{-14pt}
\fi

Thm.~\ref{thm:achievability} gives an upper bound on average blocklength $\ell$ and therefore a lower bound on achievable rate for codes with cardinality $M$ and error probability $\epsilon$.  However, it is not always straightforward to compute the achievable $(\ell, \frac{\log M}{\ell})$ pairs. In \cite[Appendix B]{Williamson_ROVA_ISIT_2013}, Williamson et al. provide a method for computing these blocklength-rate pairs, based on numerical evaluation of an infinite sum. For the AWGN channel, each term in the sum requires a 3-dimensional numerical integration. 
In \appref{sec:bound_comp} of this \paperref, we describe a different method of computing the average stopping time $\E[\tau]$ in \eqref{eqn:vlf_Etau} based on Wald's equality \cite[Ch. 5]{Gallager_stochastic_textbook_2013}. This technique is computationally simpler and does not suffer from numerical accuracy issues that arise in evaluating an infinite sum.
As explained in \appref{sec:bound_comp}, the new method applies only to channels with bounded information density $i(X_n; Y_n)$ (e.g, the BSC or BI-AWGN channel, but not the AWGN channel with real-valued inputs).

It is straightforward to extend Polyanskiy et al.'s random-coding lower bound \cite{Polyanskiy_IT_2011_NonAsym} to VLF codes derived from repeating length-$N$ mother codes, which we will show in Cor.~\ref{thm:achievability_N}. We begin by defining $(\ell, M, \epsilon)$ repeat-after-$N$ VLF codes, in which the coded symbols for $n > N$ repeat the first $N$ symbols. Let $r \in \{1, \dots, N\}$ be the index within each block of $N$ symbols, i.e., $n = s N + r$ for some $s \in \{0, 1, \dots\}$.


\begin{definition}	\label{def:vlf_N}
An ($\ell, M, \epsilon)$ repeat-after-$N$ VLF code is defined as in Def.~\ref{def:vlf}, except the following are different:
\begin{itemize}
	\item A sequence of encoder functions $f_r : {\mathcal U} \times {\mathcal W} \times {\mathcal Y}_{sN+1}^{n-1} \rightarrow 
{\mathcal X}$, which defines the $n$th channel input, where $r \in \{1, \dots, N\}$, $n \geq 1$, $s = \lfloor \frac{n-1}{N}\rfloor$:
\begin{align}
	X_n &= f_r(U, W, Y_{sN+1}^{n-1}) \, .
\end{align}

	\item A sequence of decoder functions $g_r : {\mathcal U} \times {\mathcal Y}_{sN+1}^n \rightarrow {\mathcal W}$, providing an estimate $\hat W_n$ of the message $W$, where $r \in \{1, \dots, N\}$, $n \geq 1$,  $s = \lfloor \frac{n-1}{N}\rfloor$:
\begin{align}
	\hat W_n =  g_r(U, Y_{sN+1}^n).	\label{eqn:dec_dec}
\end{align}
\end{itemize}
\end{definition}
A practical consequence of this definition is that for decision-feedback repeat-after-$N$ codes, only $N$ unique coded symbols need to be generated for each message, due to the fact that $X_n = f_r(U, W)$.
Because the decoder in \eqref{eqn:dec_dec} only uses the received symbols from the current length-$N$ block, we define the following modified information density:
\begin{align}
	i_N(X^n; Y^n) &= \log \frac{d\P(Y_{sN+1}^n = y_{sN+1}^n|X_{sN+1}^n = x_{sN+1}^n)}{d\P(Y_{sN+1}^n = y_{sN+1}^n)} \\
	&= \log \frac{d\P(Y^r = y_{sN+1}^n|X^r = x_{sN+1}^n)}{d\P(Y^r = y_{sN+1}^n)}.
\end{align}

\begin{corollary}[Random-coding lower bound for repeat-after-$N$ codes]
	Suppose that $N$ is large enough such that $\P[\tau~\leq~N]~>~0$.
	Then for a scalar $\gamma > 0$, $\exists$ an $(\ell, M, \epsilon)$ repeat-after-$N$ VLF code satisfying
	\begin{IEEEeqnarray}{rCl}
	\ell &\leq& \mathrm{E}[\tau] = \frac{\sum \limits_{n=0}^{N-1} \P[\tau > n]}{1 - \P[\tau > N]}, \\
	\epsilon &\leq& (M-1) \P[ \bar\tau \leq \tau],
	\end{IEEEeqnarray}
	where $\gamma$ is a threshold for the hitting times $\tau$ and $\bar\tau$:
	\begin{IEEEeqnarray}{rCl}
	\tau &=& \inf \{n \geq 0 :  i_N(X^n;Y^n) \geq \gamma \} 	\label{eqn:vlf_tau_N} \\
	\bar\tau &=& \inf \{n \geq 0 : i_N(\bar X^n;Y^n) \geq \gamma \}.	\label{eqn:vlf_bar_tau_N}
	\end{IEEEeqnarray}
	\label{thm:achievability_N}
\end{corollary}
\ifx \columnmode \single
   \vspace{-36pt}
\fi
Note that in \eqref{eqn:vlf_tau_N} and \eqref{eqn:vlf_bar_tau_N}, it is possible to have $n > N$.
We will show in \secref{sec:stop_feedback} that repeat-after-$N$ VLF codes constructed by puncturing convolutional codes can deliver throughput surpassing that of the random-coding lower bound of Thm.~\ref{thm:achievability}, even when the random-coding lower bound does not use the repeat-after-$N$ restriction. Similar behavior was seen in Chen et al. \cite{Chen_ISIT_2013,Chen_Feedback_Journal_2013}, which explores the effect of finite-length codewords on the achievable rates of VLFT codes. 
The proof of Cor.~\ref{thm:achievability_N} is in \appref{sec:proof_achievability_N}.

Thm.~\ref{thm:achievability_N} can also be extended to accommodate repeat-after-$N$ codes that permit decoding only at $m$ specified intervals (modulo $N_m$): $n \in \{N_1, N_2, \dots, N_m, N_m + N_1, \dots\}$. Similar to the repeat-after-$N$ setting, the coded symbols for $n > N_m$ repeat the first $N_m$ symbols. We define $I_i = N_i - N_{i-1}$ as the transmission length of the $i$th transmission ($i = 1, \dots, m$), where $N_0 = 0$ for convenience. This framework models practical systems, in which decoding is attempted after groups of symbols instead of after individual symbols. The following corollary provides the achievability result for random coding with ``packets" of length $I_i$.

\begin{corollary}[Random-coding lower bound for $m$-transmission repeat-after-$N_m$ codes]
	Suppose that $N_m$ is large enough such that $\P[\tau~\leq~N_m]~>~0$.
	Then for a scalar $\gamma > 0$, $\exists$ an $(\ell, M, \epsilon)$ $m$-transmission, repeat-after-$N_m$ VLF code satisfying 
	\begin{IEEEeqnarray}{rCl}
	\ell &\leq& \mathrm{E}[\tau] = \frac{\sum \limits_{i=0}^{m-1} I_i \P[\tau > N_i]}{1 - \P[\tau > N_m]}, \label{eqn:vlf_Etau_Nm} \\
	\epsilon &\leq& (M-1) \P[ \bar\tau \leq \tau],
	\end{IEEEeqnarray}
	where $\gamma$ is a threshold for the hitting times $\tau$ and $\bar\tau$:
\ifx \columnmode \single
	\begin{IEEEeqnarray}{rCl}
	\tau &= \inf \{&n \geq 0 : i_N(X^n;Y^n) \geq \gamma\} \cap \{N_1, N_2, \dots, N_m, N_m + N_1, \dots\}, \\
	\bar\tau &= \inf \{&n \geq 0 :  i_N(\bar X^n;Y^n) \geq \gamma\} \cap \{N_1, N_2, \dots, N_m, N_m + N_1, \dots\}.
	\end{IEEEeqnarray}
\else
	\begin{IEEEeqnarray}{rCl}
	\tau = \inf &\{& n \geq 0 : i_{N_m}(X^n;Y^n) \geq \gamma\} \nonumber \\
	 \cap &\{& N_1, N_2, \dots, N_m, N_m + N_1, \dots\}, \\
	\bar\tau = \inf &\{& n \geq 0 :  i_{N_m}(\bar X^n;Y^n) \geq \gamma\} \nonumber \\
	 \cap &\{& N_1, N_2, \dots, N_m, N_m + N_1, \dots\}.
	\end{IEEEeqnarray}
\fi
	\label{thm:achievability_Nm}
\end{corollary}
\vspace{-14pt}
The proof of Cor.~\ref{thm:achievability_Nm} is omitted. It closely follows that of Cor.~\ref{thm:achievability_N} and relies on the fact that decoding can only occur at the specified intervals, so the expected stopping time in \eqref{eqn:vlf_Etau_Nm} is a sum of probabilities weighted by the $i$th transmission length $I_i$.
%

As shown in Fig.~\ref{fig:ROVA_bsc_sims} for the binary symmetric channel (BSC) with crossover probability $p~$$=$$~0.05$, there is a considerable gap between the lower and upper bounds on the maximum rate of VLF codes at short blocklengths. Because of this gap in the fundamental limits, it is not clear what short-blocklength performance is achievable. To explore this question, we present a deterministic coding scheme in \secref{sec:stop_feedback}, fixing $M$ and $\epsilon$ to explore what rates can be achieved at short blocklengths (less than 300 symbols).

One may wonder at this point why the previously transmitted blocks of $N$ (or $N_m$) symbols are not included in the decoding.  Using these previous symbols cannot hurt and should help.  However, it helps so little that it is not worth the additional difficulty of analyzing its benefit.  Dropping earlier data is inconsequential because the probability of generating a NACK after $N$ symbols is relatively low for reasonably chosen $N$, (approximately 0.01 to 0.001 in some of our examples and even less in others). As a result, there are very few opportunities for re-using previous symbols, and the potential benefit is negligible because using more than $N$ symbols happens so rarely. 


%
%
\section{Convolutional Codes with Decision Feedback}
\label{sec:stop_feedback}

%
%
\subsection{Reliability-based Error Detection}
\label{sec:rova}
%

This section investigates the performance of punctured convolutional codes with decision feedback in the context of \secref{sec:vlf}'s VLF coding framework. Many hybrid ARQ systems with CCs use CRCs for explicit error detection at the receiver, sometimes referred to as code-based retransmission. However, at short blocklengths, the latency overhead of a CRC strong enough to meet the $\epsilon$ error constraint may result in a significant rate penalty.
Here we investigate the performance of reliability-based retransmission, in which the receiver stops decoding when the posterior probability of the decoded word is at least $(1 - \epsilon)$. This approach guarantees that the $\epsilon$ error requirement is met and does not require additional coded symbols to be sent for error detection.

For practical purposes, we consider only repeat-after-$N$ codes in this section. After receiving the $n$th transmitted symbol, the receiver determines the maximum a posteriori (MAP) message $\hat W_n$ and computes its posterior probability (reliability), where 
\begin{align}
	\hat W_n &= \arg \max \limits_{i \in {\mathcal W}} \P(W = i | Y_{sN+1}^n) \label{eqn:rova_What}.
\end{align}
The stopping rule for the reliability-based (RB) retransmission scheme is defined according to:
\begin{align}
	\tau^\text{(RB)} &= \inf \{n \geq 0 : \P(W = \hat W_n  | Y_{sN+1}^n) \geq 1 - \epsilon \}.	\label{eqn:rova_tau}
\end{align}

Finding the MAP message in \eqref{eqn:rova_What} may be accomplished by computing all $M$ posterior probabilities in \eqref{eqn:rova_What}, which can in principle be performed for any code, such as LDPC codes. However, even for moderate blocklengths, this may not be computationally feasible, similar to the complexity challenge of ML decoding. Fortunately, for terminated CCs, Raghavan and Baum's ROVA \cite{Raghavan_ROVA_TransIT_1998} gives an efficient method to compute $\P(W = \hat W_n  | Y_{sN+1}^n)$, the posterior probability of the MAP message\footnote{When the source symbols are equiprobable, there is a one-to-one correspondence between the MAP message and the ML codeword, the latter of which is identified by both the Viterbi Algorithm and the ROVA.}. 
The computational complexity of the ROVA is linear in the blocklength and exponential in the constraint length, on the same order as that of the Viterbi Algorithm.
This allows the receiver to implement the stopping rule in \eqref{eqn:rova_tau} without explicitly evaluating all $M$ posterior probabilities.
Due to this rule, the overall probability of error in the reliability-based stopping scheme will satisfy the $\epsilon$ constraint:
\begin{align}
	\P[\hat W_{\tau^\text{(RB)}} \neq W ] &= \E \big[1 - \P[\hat W_{\tau^\text{(RB)}} = W | Y_{sN+1}^{\tau^\text{(RB)}}] \big] \leq \epsilon.	\label{eqn:rova_error}
\end{align}


However, terminated CCs suffer from rate loss at short blocklengths, as described earlier, and Raghavan and Baum's ROVA \cite{Raghavan_ROVA_TransIT_1998} does not permit decoding of throughput-efficient TBCCs. Williamson et al. \cite{Williamson_TBROVA_2014}'s TB ROVA describes how to compute the posterior probability of MAP messages corresponding to tail-biting codewords. In the simulations that follow, we use the ROVA for terminated codes and, when computational complexity permits, the TB ROVA for tail-biting codes. In particular, we implement an efficient version of the TB ROVA called the Tail-Biting State-Estimation Algorithm (TB SEA) from \cite{Williamson_TBROVA_2014} that reduces the number of computations but still computes the MAP message probability exactly\footnote{The TB SEA and TB ROVA compute the same probability as long as $\P(W = \hat W_n|Y) > \frac{1}{2}$. In the proposed reliability-based retransmission scheme with $\epsilon < \frac{1}{2}$, this condition is met for $\tau^\text{(RB)} = n$, so the TB SEA is an ML sequence decoder.}.

The details of our decision-feedback scheme are as follows. Similar to \cite{Fricke_Reliability_HARQ_TCOM_2009}, 
if the computed word-error probability at blocklength $n$ is greater than the target $\epsilon$, the decoder signals that additional coded symbols are required (sends a NACK), and the transmitter sends another coded symbol. When the word-error probability is less than $\epsilon$, the decoder sends an ACK, and transmission stops.
We encode a message with $k = \log M$ message symbols into a mother codeword of length $N$. 
One symbol is transmitted at a time, using pseudo-random, rate-compatible puncturing of the mother code. At each decoding opportunity, the receiver uses all received symbols to decode and computes the MAP message probability.
If the receiver requests additional redundancy after $N$ symbols have been sent, the transmitter begins resending the original sequence of $N$ symbols and decoding starts from scratch.  
(This is a repeat-after-$N$ VLF code.)

Similar to the random-coding lower bound for repeat-after-$N$ codes in Cor.~\ref{thm:achievability_N}, we can express the latency $\lambda^\text{(RB)}$ and the throughput $R_t^\text{(RB)}$ of the proposed scheme as 
	\begin{align}
	\lambda^\text{(RB)} &\leq \frac{1 + \sum\limits_{i=1}^{N-1} P_{\text{NACK}}(i)}{1 - P_{\text{NACK}}(N)}, 
	\label{eqn:lambda_ROVA} \\
	R_t^\text{(RB)} &= \frac{k}{\lambda^\text{(RB)}} (1 - P_{\text{UE}}),
	\label{eqn:Rt_ROVA}
	\end{align}
where $P_{\text{NACK}}(i)$ is the probability that a NACK is generated because the MAP message probability is less than $(1 - \epsilon)$ when $i$  coded symbols (modulo $N$) have been received, $k$ is the number of information symbols, and $P_{\text{UE}}$ is the overall probability of undetected error. Note $P_{\text{UE}} \leq \epsilon$ by definition of the stopping rule, as shown in \eqref{eqn:rova_error}. 
We have included the factor $(1 - P_{\text{UE}})$ in the throughput expression to emphasize that we are only counting the messages that are decoded both correctly and with sufficient reliability at the receiver (i.e., the goodput). 
In \secref{sec:rova_results}, we obtain $\lambda^\text{(RB)}$, $R_t^\text{(RB)}$, and $P_\text{UE}$ empirically. See \appref{sec:sampling} for details of the estimators involved.

While some benefit can be accrued by retaining the $N$ already-transmitted symbols (for example, by Chase code combining), our analysis focuses on cases in which starting over after $N$ symbols is rare (e.g., $P_{\text{NACK}}(N) \approx 10^{-2}$ to $10^{-3}$). Thus the benefit of combining is minimal, being limited to possibly decreasing latency only in rare instances. If combining were used, the right-hand side of \eqref{eqn:lambda_ROVA} would be at least $1 + \sum\limits_{i=1}^{N-1} P_{\text{NACK}}(i)$, which is only a decrease by the multiplicative factor $(1 - P_{\text{NACK}}(N)) \approx 99\%$. For simplicity we do not exploit this opportunity in our scheme.


%
\subsection{Convolutional Code Polynomials}
\label{sec:codes}
%

\label{sec:conv_codes}

\begin{table}
\begin{center}
  \caption{Generator polynomials $g_1$, $g_2$, and $g_3$ corresponding to the rate $1$/$3$ convolutional codes used in the VLF simulations. $d_{\text{free}}$ is the free distance, $A_{d_{\text{free}}}$ is the number of codewords with weight $d_{\text{free}}$, and $L_D$ is the analytic traceback depth.} 
\begin{tabular}{ c | c | c | c | c | c }
  \# Memory  & \# States & Polynomial & & & \\
  Elements, $\nu$ & $s=2^{\nu}$ & $(g_1, g_2, g_3)$  & $d_{\text{free}}$ & $A_{d_{\text {free}}}$ & $L_D$ \\
  \hline \hline
  6 & 64 & (117, 127, 155) & 15 & 3 & 21 \\
  8 & 256 & (575, 623, 727) & 18 & 1 & 25 \\
  10 & 1024 & (2325, 2731, 3747) & 22 & 7 & 34 \\
  \hline
\end{tabular}
\label{tbl:vlf_conv_codes}
\end{center}
\end{table}


This section briefly lists the convolutional code polynomials used in the subsequent VLF coding simulations. 
We use both terminated convolutional codes and tail-biting convolutional codes for the ROVA-based stopping rule. Comparisons with CRCs use only tail-biting convolutional codes.

Table \ref{tbl:vlf_conv_codes}, taken from Lin and Costello \cite[Table 12.1]{Lin_2004_ECC}, lists the generator polynomials for the rate-$1$/$3$ convolutional codes that were used as the mother codes for our simulations. Each code selected has the optimum free distance $d_{\text{free}}$, which is listed along with the analytic traceback depth $L_D$ \cite{Anderson_Traceback_TransIT_1989}. 
Higher-rate codewords used for the incremental transmissions are created by pseudorandom, rate-compatible puncturing of the rate-$1$/$3$ mother codes.

All of the simulations involving AWGN use the BI-AWGN channel (i.e., using BPSK signaling) with soft-decision decoding.
The BI-AWGN channel has a maximum Shannon capacity of 1 bit per channel use, even when the SNR $\eta$ is unbounded. We have included comparisons with the asymptotic capacity of the BI-AWGN channel as well as the full AWGN channel (i.e., with real-valued inputs drawn i.i.d. $\sim \mathcal{N}(0,\eta)$).

%
\subsection{Numerical Results}
\label{sec:rova_results}

\begin{figure}
  \centering
\ifx \columnmode \single
    \scalebox{0.7}{\includegraphics{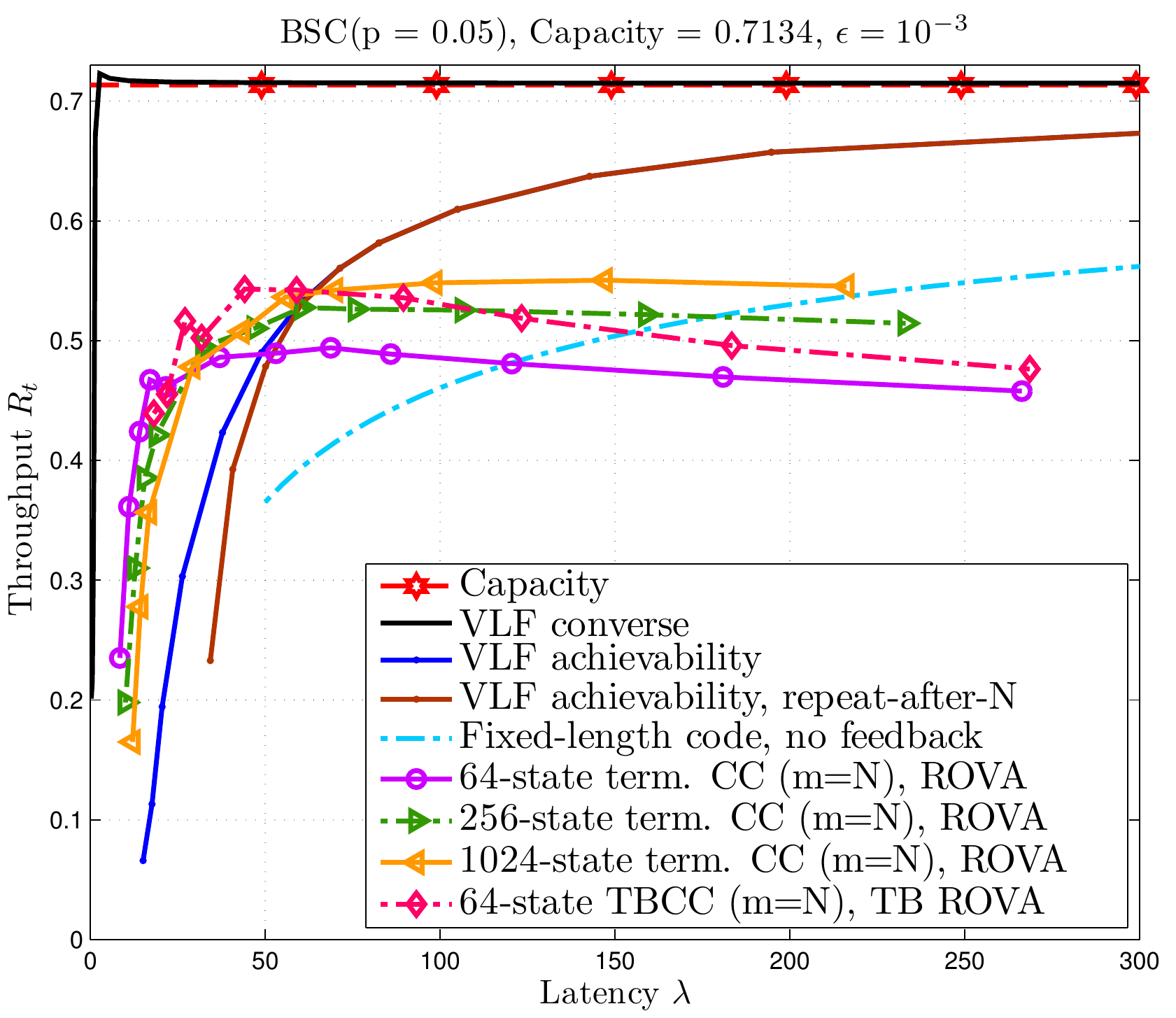}}
\else
    \scalebox{0.5}{\includegraphics{willi2.pdf}}
\fi
  \caption{Short-blocklength performance of the reliability-based retransmission scheme over the BSC$(p=0.05)$ with target probability of error $\epsilon$~=~$10^{-3}$.
Simulations use the ROVA for terminated convolutional codes (term. CC) and the TB ROVA for tail-biting convolutional codes (TBCC), with decoding after every symbol ($m$$=$$N$).}
   \label{fig:ROVA_bsc_sims}
\ifx \thesismode \journal
\fi
\end{figure}

Fig.~\ref{fig:ROVA_bsc_sims} illustrates the short-blocklength performance of the reliability-based retransmission scheme using the ROVA for terminated CCs (term. CC) and the TB ROVA for TBCCs, compared to the fundamental limits for VLF codes. This example uses the BSC with crossover probability $p$$=$$0.05$ and target probability of error $\epsilon$$=$$10^{-3}$.
The points on each CC curve correspond to different values of the information length $k$. In general, the latency increases with $k$.
The Shannon (asymptotic) capacity of the BSC with crossover probability $p$ is $C_\text{BSC}$~=~$1-h_\text{b}(p)$. 
The random-coding lower bound (`VLF achievability') is from Thm.~\ref{thm:achievability} and the upper bound (`VLF converse') is from \convbsc. 
An example of the random-coding lower bound for repeat-after-$N$ codes from Cor.~\ref{thm:achievability_N} is also shown (`VLF achievability, repeat-after-N'), with $N = 3 \log M$, which corresponds to our implementations with a rate-1/3 mother code.
Both the convolutional code simulations and the VLF bounds correspond to decoding after every received symbol. 
Fig. \ref{fig:ROVA_bsc_sims} also includes the maximum rate at finite blocklengths without feedback (`Fixed-length code, no feedback'), based on the normal approximation from \cite{Polyanskiy_CCR_2010}.

Though the upper and lower bounds for VLF codes coincide asymptotically, there is a considerable gap when latency is below 100 bits, a region in which convolutional codes can deliver good performance. At the shortest blocklengths, the 64-state code with the fewest memory elements performs best among the terminated codes, due to the increased rate loss of the codes with larger constraint lengths. However, as the message length $k$ increases (and the latency increases), the more powerful 1024-state terminated code delivers superior throughput. As the latency continues to increase, the codes' throughputs fall below that of the VLF achievability bound, which is based on random coding.  Random coding improves with latency, but the word-error probability of convolutional codes increases with blocklength once the blocklength is beyond twice the traceback depth $L_D$ of the convolutional code \cite{Anderson_Traceback_TransIT_1989}.

The maximum throughput obtained for the BSC simulations in Fig.~\ref{fig:ROVA_bsc_sims} is $R_t^{(\text{RB})} = 0.551$ bits per channel use at $\lambda^{(\text{RB})} = 147.1$ bits for the $k=91$, 1024-state terminated CC, which is $77.2\%$ of the BSC capacity. However, using TBCCs allows codes with fewer memory elements to achieve similar rates at much lower latency. The maximum throughput for the 64-state TBCC  is $R_t^{(\text{RB})} = 0.543$ bits per channel use at $\lambda^{(\text{RB})} = 44.1$ bits for the $k=24$, 64-state TBCC.
%
Note the curves for both terminated and tail-biting 64-state CCs exhibit non-monotonic behavior (near $\lambda^\text{(RB)}$$=$$17$, $k$$=$$8$ and $\lambda^\text{(RB)}$$=$$27$, $k$$=$$14$, respectively). This is likely due to non-monotonic minimum distance growth of the terminated convolutional codes as a function of blocklength, in conjunction with non-ideal effects of pseudo-random puncturing.
%

\begin{figure}
  \centering
\ifx \columnmode \single
    \scalebox{0.7}{\includegraphics{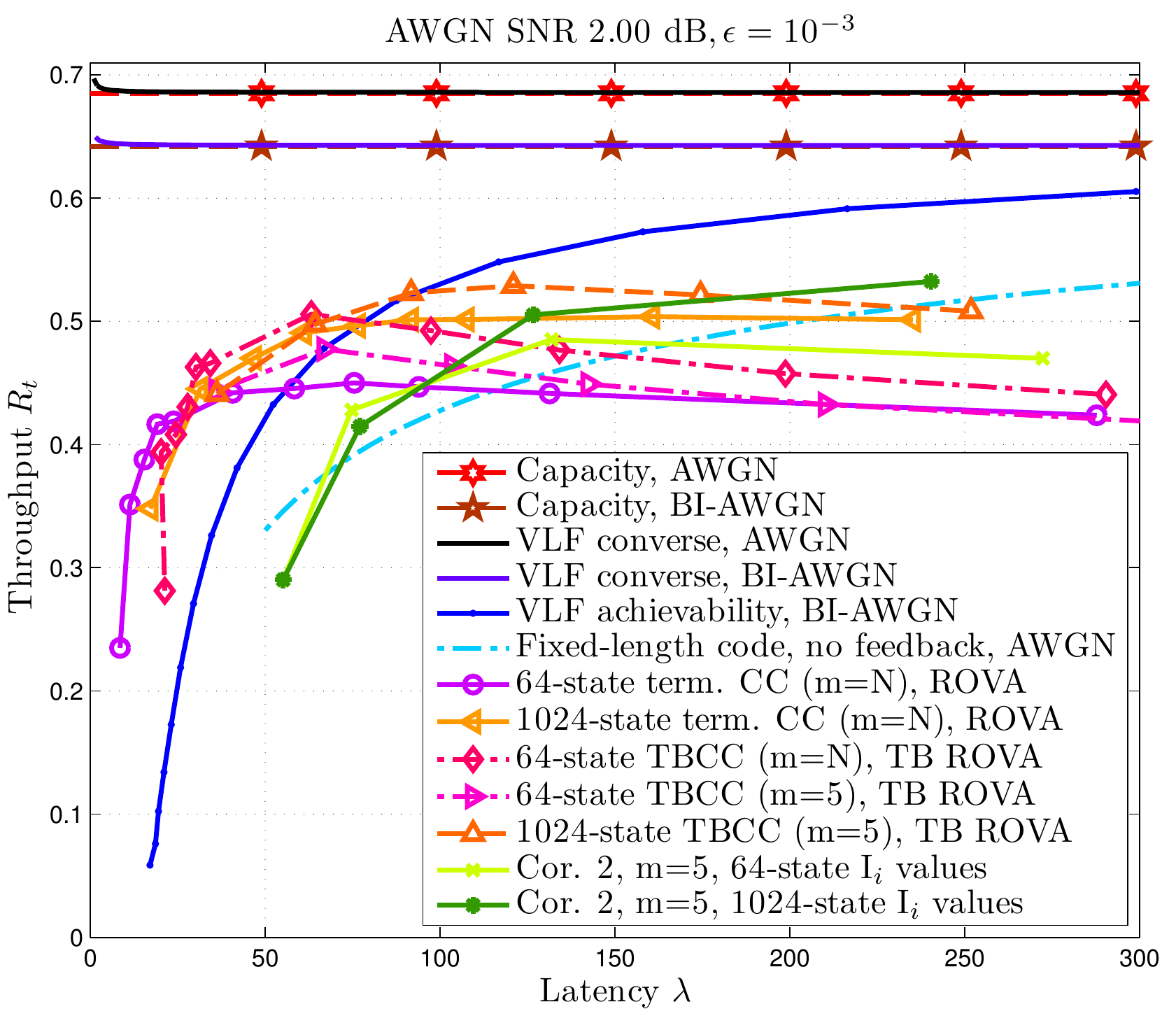}}
\else
    \scalebox{0.5}{\includegraphics{willi3.pdf}}
\fi
  \caption{Short-blocklength performance of the reliability-based retransmission scheme over the AWGN channel with SNR 2.00 dB and target probability of error $\epsilon$~=~$10^{-3}$.
Simulations use the ROVA for terminated convolutional codes (term. CC) and the TB ROVA for tail-biting convolutional codes (TBCC), with decoding after every symbol ($m$$=$$N$) or with decoding after $m$$=$$5$ groups of symbols.}
   \label {fig:ROVA_awgn_sims}
\ifx \thesismode \journal
\fi
\end{figure}


Fig.~\ref{fig:ROVA_awgn_sims} shows the performance of the reliability-based retransmission scheme over the AWGN channel with SNR 2 dB and target $\epsilon$$=$$10^{-3}$. Fig.~\ref{fig:ROVA_awgn_sims} includes results from the ROVA for terminated CCs and the TB ROVA for TBCCs.
The Shannon capacity of the AWGN channel with SNR $\eta$ is $C_\text{AWGN} = \frac{1}{2}\log (1 + \eta)$, and the Shannon capacity of the BI-AWGN channel is approximated as in \cite{Nasif_BIAWGN_Capacity_CISS_2005}.
The random-coding lower bound (`VLF achievability') is from Thm.~\ref{thm:achievability}, the AWGN upper bound (`VLF converse, AWGN') is from \convawgn  ~ and the BI-AWGN upper bound (`VLF converse, BI-AWGN') is from \convbsc, both particularized to the Gaussian channel.
The ``Fixed-length code, no feedback"  curve uses the normal approximation \cite{Polyanskiy_CCR_2010}.
The CC simulations in Fig.~\ref{fig:ROVA_awgn_sims} deliver similar performance to those in the BSC case of Fig.~\ref{fig:ROVA_bsc_sims}. The throughput of the convolutional codes surpasses the random-coding lower bound at short blocklengths, but plateaus around latencies of 100 bits.

Convolutional codes with more memory elements would be expected to deliver improved throughput, but computational complexity limits us to codes with 1024 states or fewer.
In both Figs.~\ref{fig:ROVA_bsc_sims} and \ref{fig:ROVA_awgn_sims}, the high decoding complexity of the 1024-state codes prevented us from using the TB ROVA when decoding after every symbol.

%
\subsection{Decoding after Groups of Symbols}
\label{sec:increments}

Decoding less frequently is practically desirable due to the round-trip delay inherent in the feedback loop and because of the complexity associated with performing the ROVA after each received symbol. Decoding with the ROVA only after packets is a natural extension of the proposed scheme, akin to the $m$-transmission repeat-after-$N_m$ codes  in \secref{sec:vlf}.
When decoding only after packets are received, the latency $\lambda^\text{(RB)}$ and the throughput $R_t^\text{(RB)}$ become
	\begin{align}
	\lambda^\text{(RB)} &\leq \frac{I_1 + \sum\limits_{i=1}^{m-1} I_i P_{\text{NACK}}(N_i)}{1 - P_{\text{NACK}}(N_m)}, 
	\label{eqn:lambda_ROVA_Nm} \\
	R_t^\text{(RB)} &= \frac{k}{\lambda^\text{(RB)}} (1 - P_{\text{UE}}).
	\label{eqn:Rt_ROVA_Nm}
	\end{align}
Here $P_{\text{NACK}}(N_i)$ is the probability of retransmission when $N_i$  coded symbols have been received. The incremental transmission length at transmission $i$ is $I_i$ and the cumulative decoding blocklength is $N_i = I_1 + \dots + I_i$.

A main challenge in an $m$-transmission incremental redundancy scheme is to select the set of $m$ incremental transmission lengths $\{I_i\}_{i=1}^m$ that provide the best rate at short blocklengths. In general, latency (resp., throughput) expressions such as \eqref{eqn:lambda_ROVA_Nm} (resp., \eqref{eqn:Rt_ROVA_Nm}) are not convex (resp., concave) in the blocklengths and must be optimized numerically.
\appref{sec:blocklength_algo} presents an algorithm to optimize the blocklengths in general incremental redundancy schemes. 
\appref{sec:algo_tb_rova} describes how to particularize the algorithm in order to select the $m$$=$$5$ optimal blocklengths in the reliability-based retransmission scheme using the TB ROVA for TBCCs.
We have chosen to evaluate the performance of this scheme with $m$$=$$5$ because increasing $m$ further brings diminishing returns. See, e.g., Chen et al. \cite{Chen_Feedback_Journal_2013}, for an examination of the impact of $m$ on the performance of VLFT codes at short blocklengths.

Table~\ref{tbl:opt_blocklengths_tb_rova} shows the optimal transmission lengths identified by the blocklength-selection algorithm. Based on these blocklengths, we simulated the TB ROVA with TBCCs in an $m$$=$$5$ transmission decision-feedback scheme.
 Fig.~\ref{fig:ROVA_awgn_sims} shows the impact on throughput when decoding is limited to these specified decoding opportunities. Despite fewer opportunities for decoding (and hence fewer chances to stop transmission early), both the 64-state and 1024-state tail-biting codes in the optimized $m$$=$$5$ setting deliver excellent performance compared to the respective terminated codes that allow decoding after every symbol (i.e., $m$$=$$N$).
Note also how at blocklengths less than approximately 75 bits, the $m$$=$$5$ TBCCs deliver higher rates than the random-coding lower bound that requires decoding after every symbol (`VLF achievability, BI-AWGN'). When compared to Cor.~\ref{thm:achievability_Nm}'s random-coding lower bound for repeat-after-$N_m$ codes on the AWGN channel, the $m$$=$$5$ TBCCs deliver higher rates for blocklengths up to about 125 bits. The `Cor. 2, m=5, 64-state I$_i$ values' curve uses the optimal $m$$=$$5$ blocklengths for the 64-state TBCC, and the 1024-state curve uses the optimal $m$$=$$5$ blocklengths for the 1024-state TBCC.
The maximum throughput obtained from these  $m$$=$$5$ simulations is $R_t^{(\text{RB})} = 0.529$ bits per channel use at $\lambda^{(\text{RB})} = 121.0$ bits, for the $k$$=$$64$, 1024-state TBCC. This is $77.2\%$ of the AWGN capacity and $82.4\%$ of the BI-AWGN capacity.

For all of the VLF simulations in this section, at least 25 undetected word-errors were accumulated for each value of $k$. Because the ROVA-based stopping rule with target $\epsilon$$=$$10^{-3}$ guarantees that the average probability of error is no more than $10^{-3}$, our simulations are not intended to estimate error rate, but rather to accurately estimate throughput and latency.  For these estimates, our simulations with 25 errors provide acceptable accuracy. 
 \appref{sec:sampling} provides an explanation for why VLF simulations with at least 25 word errors are sufficient for reliably estimating the throughput and latency. 
Additionally, many of the simulations did in fact have more than 100 word errors accumulated, including the $m$$=$$5$ TBCCs listed in Table~\ref{tbl:opt_blocklengths_tb_rova}. Tallying at least 100 errors roughly corresponds to a confidence interval of $\pm0.2  P_\text{UE}$ with confidence level $95\%$, whereas tallying only 25 errors corresponds to a confidence interval of $\pm0.4  P_\text{UE}$ with confidence level $95\%$, as discussed in \appref{sec:sampling}.


\begin{table}[t] 
\begin{center}
  \caption{Optimal transmission lengths $\{I_i\}^*$ for the $m$$=$$5$ transmission scheme using the TB ROVA, for SNR $\eta$$=$$2$ dB, along with the simulated error probability $P_\text{UE}$ corresponding to target error probability $\epsilon$$=$$10^{-3}$.} \label{tbl:Ivalues_tb_rova}
\ifx \columnmode \single
\begin{tabular}{ c | c | c | c | c | c }
  &  & \multicolumn{2}{|c|}{64-state TBCC}    & \multicolumn{2}{|c}{1024-state TBCC}  \\
  Info. Bits & Target & Transmission Lengths & Simulated   	& Transmission Lengths & Simulated  \\
  $k$ &  Error $\epsilon$ &  $\{I_1^*, I_2^*, I_3^*, I_4^*, I_5^*\}$ &  Error $P_{\text{UE}}$ 	&  $\{I_1^*, I_2^*, I_3^*, I_4^*, I_5^*\}$ & Error  $P_{\text{UE}}$ \\
  \hline \hline
	16 	& $10^{-3}$ & 30, 	3, 	3, 	5, 	7  & $2.26 \times 10^{-4}$ 	& 	29, 	4, 	4, 	4, 	7  & $2.47 \times 10^{-4}$  \\
	32 	& $10^{-3}$ & 57, 	6, 	7, 	9, 	16 & $1.96 \times 10^{-4}$ 	& 	56, 	5, 	5, 	7, 	12 & $1.98 \times 10^{-4}$ \\
	48 	& $10^{-3}$ & 88, 	9, 	10, 	13, 	24 & $2.35 \times 10^{-4}$ 	&	80, 	7, 	7, 	9, 	16 & $2.09 \times 10^{-4}$ \\
	64	& $10^{-3}$ & 121, 	12, 	13, 	17, 	29 & $2.60 \times 10^{-4}$  &		106, 	9, 	9, 	12, 	22 & $1.99 \times 10^{-4}$ \\
	91 	& $10^{-3}$ & 178, 17, 	18, 	22, 	38 & $2.65 \times 10^{-4}$ 	& 	151, 	13, 	14, 	17, 	31 & $2.20 \times 10^{-4}$ \\
	128 	& $10^{-3}$ & 261, 	23, 	24, 	30, 	46 & $2.44 \times 10^{-4}$ & 	223, 	17, 	18, 	24, 	44 & $2.34 \times 10^{-4}$ \\
  \hline
\end{tabular}
\else
\begin{tabular}{ c | c | c | c }
  &  & \multicolumn{2}{c}{64-state TBCC}    \\
  Info. Bits & Target & Transmission Lengths & Simulated   	\\
  $k$ &  Error $\epsilon$ &  $\{I_1^*, I_2^*, I_3^*, I_4^*, I_5^*\}$ &  Error $P_{\text{UE}}$  \\
  \hline 
	16 	& $10^{-3}$ & 30, 	3, 	3, 	5, 	7  & $2.260 \times 10^{-4}$ 	\\
	32 	& $10^{-3}$ & 57, 	6, 	7, 	9, 	16 & $1.960 \times 10^{-4}$ 	\\
	48 	& $10^{-3}$ & 88, 	9, 	10, 	13, 	24 & $2.350 \times 10^{-4}$ 	\\
	64	& $10^{-3}$ & 121, 	12, 	13, 	17, 	29 & $2.600 \times 10^{-4}$  \\
	91 	& $10^{-3}$ & 178, 17, 	18, 	22, 	38 & $2.650 \times 10^{-4}$ 	 \\
	128 	& $10^{-3}$ & 261, 	23, 	24, 	30, 	46 & $2.440 \times 10^{-4}$ \\
\multicolumn{4}{c}{} \\
  &  & \multicolumn{2}{c}{1024-state TBCC}    \\
  Info. Bits & Target & Transmission Lengths & Simulated   	\\
  $k$ &  Error $\epsilon$ &  $\{I_1^*, I_2^*, I_3^*, I_4^*, I_5^*\}$ &  Error $P_{\text{UE}}$  \\
  \hline 
	16 	& $10^{-3}$ & 29, 	4, 	4, 	4, 	7  & $2.473 \times 10^{-4}$  \\
	32 	& $10^{-3}$ & 56, 	5, 	5, 	7, 	12 & $1.976 \times 10^{-4}$ \\
	48 	& $10^{-3}$ & 80, 	7, 	7, 	9, 	16 & $2.085 \times 10^{-4}$ \\
	64	& $10^{-3}$ & 106, 	9, 	9, 	12, 	22 & $1.993 \times 10^{-4}$ \\
	91 	& $10^{-3}$ & 151, 	13, 	14, 	17, 	31 & $2.197 \times 10^{-4}$ \\
	128 	& $10^{-3}$ & 223, 	17, 	18, 	24, 	44 & $2.344 \times 10^{-4}$ \\
\end{tabular}
\fi
\label{tbl:opt_blocklengths_tb_rova}
\end{center}
\end{table}

%
%
\subsection{Performance for Different SNRs}

In Figs.~\ref{fig:ROVA_bsc_sims} and \ref{fig:ROVA_awgn_sims} above, the blocklengths were optimized for one specific parameter set: $\epsilon$$=$$10^{-3}$, SNR 2 dB, fixed $k$, and a particular generator polynomial. Adapting these codes to channels with different SNRs or error requirements would require extensive re-characterization of the retransmission probabilities and optimization of the blocklengths. However, it is possible instead to use a heuristic choice of the $m$ blocklengths that provides good throughput (if not optimal) across a limited range of SNRs. 

\begin{figure}
  \centering
\ifx \columnmode \single
    \scalebox{0.7}{\includegraphics{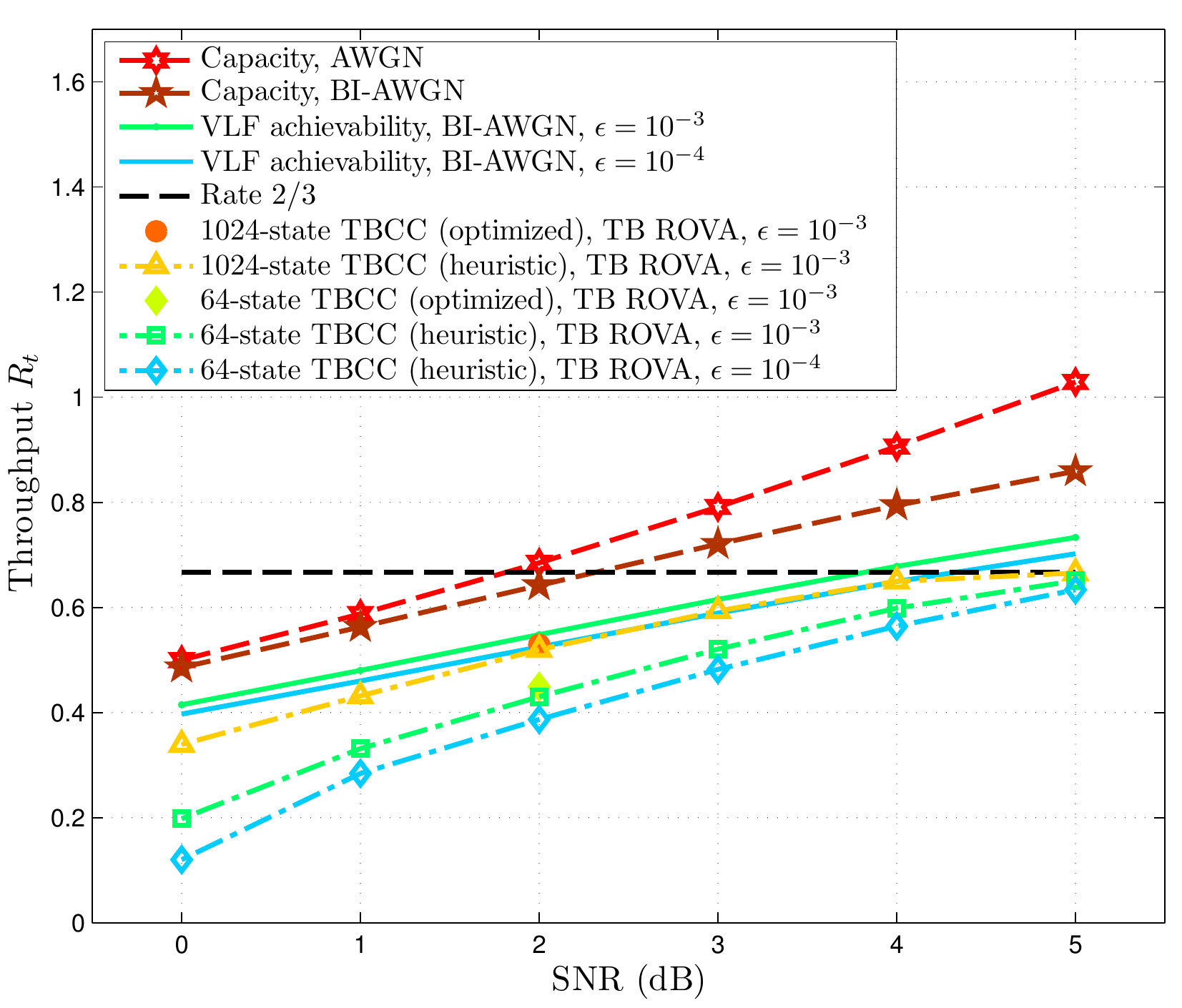}}
\else
    \scalebox{0.5}{\includegraphics{willi4.pdf}}
\fi
    \hspace{0.0in}
  \caption{Performance of a heuristic blocklength-selection policy for $\{I_i\}_{i=1}^{5}$ used across a range of SNRs for 64-state and 1024-state tail-biting convolutional codes (TBCC) with $k$$=$$64$ message bits. Two different error targets $\epsilon$ are evaluated.}
   \label{fig:ROVA_AWGN_SNR_range}
\end{figure}

Fig.~\ref{fig:ROVA_AWGN_SNR_range} shows an example of 64-state and 1024-state TBCCs simulated across a range of SNRs with the same heuristic choice of $m$$=$$5$ blocklengths. In particular, for $k$$=$$64$, the $m$$=$$5$ blocklengths were $\{ I_1$$=$$\frac{3}{2}k$$=$$96, I_2$$=$$\frac{1}{4} k$$=$$16, I_3$$=$$\frac{1}{4} k$$=$$16, I_4$$=$$\frac{1}{4} k$$=$$16, I_5$$=$$\frac{3}{4} k$$=$$48\}$.
The corresponding 2 dB throughput for the optimized $m$$=$$5$, $k$$=$$64$ blocklengths from Fig.~\ref{fig:ROVA_awgn_sims} is also shown. For example, the throughput corresponding to the optimized blocklengths for the 1024-state code is $0.529$, whereas the throughput for the heuristic blocklengths is $0.520$. Fig.~\ref{fig:ROVA_AWGN_SNR_range} demonstrates that as the SNR increases, the maximum rate achieved by this particular heuristic blocklength-selection policy is $\frac{2}{3}$, since the highest rate possible is $\frac{k}{I_1}$$=$$\frac{2}{3}$. This example shows that a more aggressive (higher rate) initial length $I_1$ should be chosen if SNRs above 4 dB are expected, which may reduce the throughput at low SNRs.

Also shown in Fig.~\ref{fig:ROVA_AWGN_SNR_range} is the performance corresponding to a lower target error constraint of $\epsilon$$=$$10^{-4}$ in addition to the original constraint of $\epsilon$$=$$10^{-3}$. Note that the actual probability of undetected error may be significantly lower. For example, for the 1024-state TBCC with $\epsilon$$=$$10^{-3}$ at SNR 5 dB, $P_\text{UE}$$<$$10^{-6}$.

%
\subsection{Peak Latency Constraints}
\label{sec:no_repeat}

Suppose that in order to meet  a peak latency constraint, instead of starting over after $m$ failed transmissions, the receiver declares an error after the $m$th failed decoding attempt. That is, when $\P(W = \hat W_{N_m}  | Y_1^{N_m}) < 1 - \epsilon$, the transmitter ignores the unreliable tentative decision $\hat W_{N_m}$ and moves on to the next message.
Whether this modified scheme meets the $\epsilon$ error requirement depends crucially on $P_{\text{NACK}}(N_m)$ compared to $\epsilon$.
For example, for all of the 1024-state TBCCs shown in Table~\ref{tbl:opt_blocklengths_tb_rova} except the TBCC with $k$$=$$16$ information bits, $P_{\text{NACK}}(N_m) \approx 4 \times 10^{-3} > \epsilon$. For the $k$$=$$16$ TBCC, $P_{\text{NACK}}(N_m) \approx 1 \times 10^{-2}$. These codes would not meet the $\epsilon$ error requirement.

There are two types of peak latency constraints: a constraint on the maximum number of transmissions $m$ and a constraint on the the maximum number of transmitted symbols $N_m$. 
When the constraint is on the number of transmissions $m$, the final transmission length $I_m$ can be increased so that $P_{\text{NACK}}(N_m) < \epsilon$, at the expense of increased latency. The blocklength-selection algorithm in \appref{sec:blocklength_algo} can be modified to satisfy this constraint. (In our examples with rate-$1/3$ TBCCs, the maximum possible blocklength would be $N_m = 3 k$.)

\begin{figure}
  \centering
\ifx \columnmode \single
    \scalebox{0.5}{\includegraphics{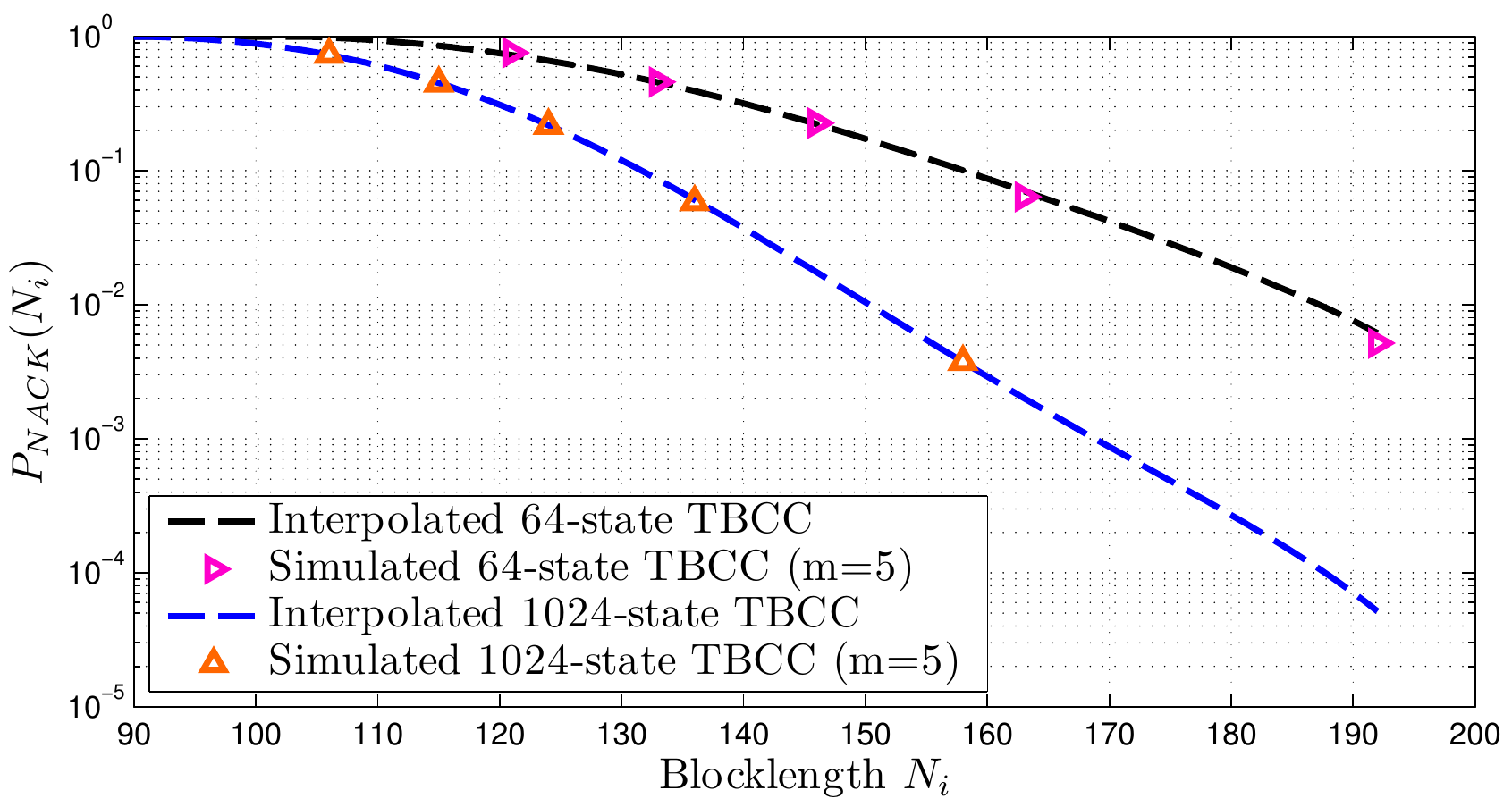}}
\else
    \scalebox{0.5}{\includegraphics{willi5.pdf}}
\fi
    \hspace{0.0in}
  \caption{$P_\text{NACK}(N_i)$ for the TBCCs from Table~\ref{tbl:opt_blocklengths_tb_rova} with $k$$=$$64$ information bits, for $i = 1, \dots, 5$. The simulated blocklengths were optimized for an error target of $\epsilon$$=$$10^{-3}$.}
   \label{fig:retrans_prob}
\end{figure}

However, if the constraint is on the maximum number of transmitted symbols $N_m$, then it may not be possible for a given mother code to meet the $\epsilon$ error requirement.
For example, Fig.~\ref{fig:retrans_prob} shows the probabilities $P_\text{NACK}(N_i)$ for the $k$$=$$64$ TBCCs from Table~\ref{tbl:opt_blocklengths_tb_rova}. Also shown are the estimated probabilities $P_\text{NACK}(N)$ obtained from interpolation, described in \appref{sec:algo_tb_rova}. If the constraint is $N_m \leq 180$, then the 64-state TBCC cannot be expected to meet the $\epsilon$$=$$10^{-3}$ requirement, whereas for the 1024-state TBCC, $N_m$ can be chosen large enough to meet the error requirement.  

%
\subsection{Code-based Error Detection}
\label{sec:CRCs}

In practice, decision-feedback schemes often use a checksum (e.g., a CRC) at the receiver to detect errors in the decoded word. However, the additional parity bits of a CRC that must be sent impose a latency cost that may be severe at short blocklengths. 
For an $A$-bit CRC appended to $k$ message bits, the throughput (not counting the check bits as information) is  
$R_{t}^{\text{(CRC)}} = \frac{k}{k+A} R_{t}$, where $R_t$ is defined similarly to \eqref{eqn:Rt_ROVA} for the reliability-based scheme. Equivalently, the rate-loss factor from an $A$-bit CRC is $\frac{A}{k+A}$. 
Using an error-detection code to determine retransmission requests is sometimes referred to as code-based error detection \cite{Fricke_Reliability_HARQ_TCOM_2009}, in contrast to reliability-based error detection with the ROVA. 
As noted in Frick and Hoeher's \cite{Fricke_Reliability_HARQ_TCOM_2009} investigation of reliability-based hybrid ARQ schemes, the rate loss and undetected error probability of the code-based approach depend critically on the blocklengths and target error probabilities involved.

\begin{figure}
  \centering
\ifx \columnmode \single
    \scalebox{0.7}{\includegraphics{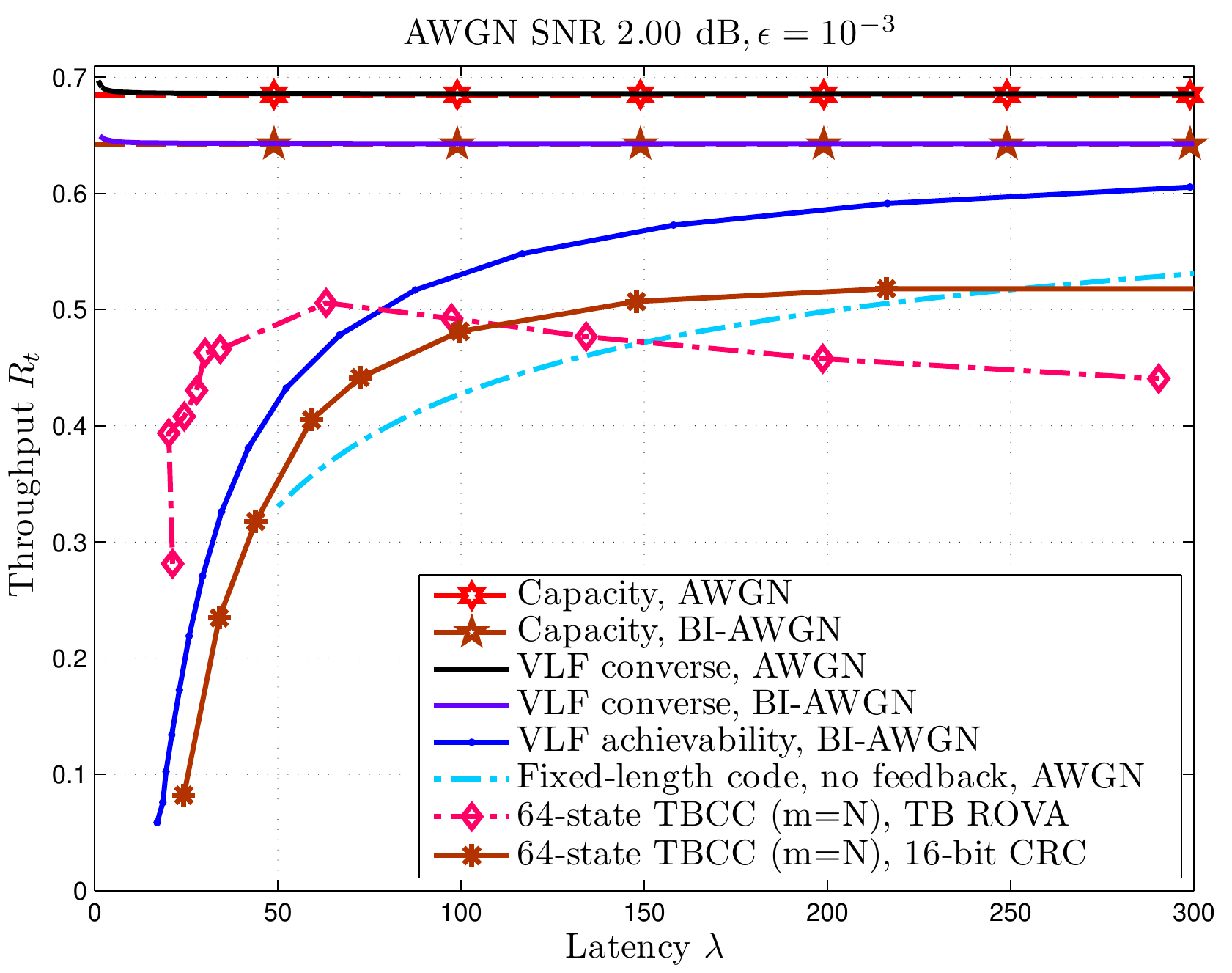}}
\else
    \scalebox{0.5}{\includegraphics{willi6.pdf}}
\fi
    \hspace{0.0in}
  \caption{A comparison of two decision-feedback schemes, the reliability-based retransmission scheme with the TB  ROVA and the code-based approach with 16-bit CRCs. Using the TB ROVA guarantees the probability of error to be less than $\epsilon$, but CRCs provide no such guarantee. In this example, 16 CRC bits were sufficient to meet the target of $\epsilon$$=$$10^{-3}$.}
   \label{fig:ROVA_CRCs}
\ifx \columnmode \single
   \vspace{-30pt}
\fi
\end{figure}

Fig.~\ref{fig:ROVA_CRCs} provides an example of the throughput obtained when decoding after every symbol and using a 16-bit CRC for error detection. 
After decoding the 64-state TBCC (which has $k+A$ input bits), the receiver re-computes the CRC to check for errors.
As expected, the rate loss of the CRCs at blocklengths less than 50 bits severely limits the achievable rates. The 64-state TBCCs decoded with the TB ROVA  deliver higher rates until the average blocklength reaches about 75 to 100 bits.
For moderately large blocklengths (e.g., 150 bits and greater), the throughput penalty induced by CRCs becomes less severe. (As the information length $k$ increases, the rate-loss factor $\frac{A}{k+A}$ decays to zero.)
Note that decoding after every symbol prevents simulation of higher-constraint-length convolutional codes (e.g., 1024-state codes).

\begin{table}
\begin{center}
  \caption{Simulated error probabilities $P_{\text{UE}}$ of the 16-bit CRC simulations in Fig.~\ref{fig:ROVA_CRCs} for the 2 dB AWGN channel. The 12-bit CRCs fail to meet the error constraint of $\epsilon$$=$$10^{-3}$, but the stronger 16-bit CRCs generally satisfy the constraint. Generator polynomials for ``good" $A$-bit CRCs from \cite{Koopman_CRC_sel_2004} are listed in hexadecimal (e.g., 0xcd indicates the polynomial is $x^8 + x^7 + x^4 + x^3 + x + 1$). The $k+A$ column indicates the number of input bits to the rate-$1/3$ convolutional encoder.
Cells labeled $-$ indicate that no simulations were run for that value of $k+A$.} 
\begin{tabular}{ c | c | c}
  Input Bits  &  $P_{\text{UE}}$  & $P_{\text{UE}}$ \\
  $k+A$ & 12-bit CRC (0xc07) & 16-bit CRC (0x8810) \\
  \hline \hline
  24 & $1.479 \times 10^{-1}$ & $1.000 \times 10^{-4}$ \\
  30 & $-$ & $8.618 \times 10^{-4}$ \\
  34 & $1.036 \times 10^{-3}$ & $1.077 \times 10^{-3}$ \\
  40 & $-$ & $2.105 \times 10^{-4}$ \\
  48 & $2.935 \times 10^{-3}$ & $2.025 \times 10^{-4}$ \\
  64 & $3.504 \times 10^{-3}$ & $2.538 \times 10^{-4}$ \\
  91 & $4.646 \times 10^{-3}$ & $3.017 \times 10^{-4}$ \\
 128 & $6.755 \times 10^{-3}$ & $5.370 \times 10^{-4}$ \\
 181 & $8.864 \times 10^{-3}$ & $6.667 \times 10^{-4}$ \\
  \hline
\end{tabular}
\label{tbl:CRC_errors}
\end{center}
\ifx \thesismode \journal
\ifx \columnmode \single
   \vspace{-30pt}
\fi
\fi
\end{table}

Importantly, using ROVA guarantees the probability of error to be less than $\epsilon$ (as long as the length-$N$ mother code is long enough to meet this constraint on average), but CRCs provide no such guarantee. In this example, in fact, TBCC simulations with 12-bit CRCs failed to meet the target of $\epsilon$$=$$10^{-3}$, as shown in Table~\ref{tbl:CRC_errors}. As a result, the codes with 12-bit CRCs do not qualify as $(\ell, M$$=$$2^k, \epsilon$$=$$10^{-3})$ VLF codes, so the VLF codes with 12-bit CRCs are not plotted in Fig.~\ref{fig:ROVA_CRCs}. Both the 12-bit and 16-bit CRC polynomials are from \cite{Koopman_CRC_sel_2004} and are listed in Table~\ref{tbl:CRC_errors}.

The 16-bit CRCs in this example generally provide sufficiently low error probabilities, but at the expense of reduced rate versus the 12-bit CRCs. One exception when the 16-bit CRC fails to meet the $\epsilon$$=$$10^{-3}$ constraint in our simulations is when $k+A=34$ ($k \approx A = 16$), as shown in Table~\ref{tbl:CRC_errors}. This high error probability seems to be an outlier compared to the other 16-bit CRC simulations, but is is consistent with  findings from previous CRC research, such as \cite{Witzke_Leung_TCOM_CRC_1985}. In \cite{Witzke_Leung_TCOM_CRC_1985}, Witzke and Leung show that the undetected error probability of a given $A$-bit CRC polynomial over the BSC can vary widely as a function of the channel bit-error rate, especially for small values of $k$. The present variable-length simulations complicate matters further, due to the stopping rule that terminates when the CRC matches, even if erroneously. Additional simulations with $k+A = 30$ and $k+A = 40$ were performed in order to illustrate the sensitivity of the error probability to the information length  $k$ (Table~\ref{tbl:CRC_errors}).

In general, it is difficult to determine exactly how many CRC bits should be used to provide the maximum throughput for hybrid ARQ schemes, since the error probability depends on the SNR. Communication system designers will likely be tempted to conservatively select large CRC lengths, which restricts the short-blocklength throughput. In contrast, the ROVA-based approach always guarantees a maximum error probability.
Still, future work that explores the performance of VLF codes at larger blocklengths (e.g., 400 bits and above) may benefit from code-based error detection.
Another challenge is that the error probability when using CRCs depends on the the underlying CC, and treating the inputs to the CRC as having passed through a BSC leads to sub-optimal designs. The recent work by Lou et al. \cite{Lou_CRC_Design} provides analytical methods for evaluating the undetected-error probability of joint CRC/CC systems and show how to select the optimal CRC polynomial for a given CC. These methods may be helpful for designing low-latency VLF codes using CRCs.

%
%
\section{Conclusion}
\label{sec:vlf_conc}
%

This \paperref~demonstrated a reliability-based decision-feedback scheme that provides throughput surpassing the random-coding lower bound at short blocklengths. We selected convolutional codes for their excellent performance at short blocklengths and used the tail-biting ROVA to avoid the rate loss of terminated convolutional codes. For both the BSC and AWGN channels, convolutional codes provided throughput above $77\%$ of capacity, with blocklengths less than 150 bits.
While codes with higher constraint lengths would be expected to provide superior performance, computational considerations limited us to evaluate 64-state codes with the TB ROVA and decoding after every symbol, and 1024-state codes with the TB ROVA in an $m$$=$$5$ transmission incremental redundancy setting. We introduced a novel blocklength-selection algorithm to aid in selecting the $m$$=$$5$ transmission lengths and showed that despite the limitations on decoding frequency, the incremental redundancy scheme is competitive with decoding after every symbol. Finally, we demonstrated that the latency overhead of CRCs imposes a severe rate-loss penalty at short blocklengths, whereas reliability-based decoding does not require transmission of separate error-detection bits.

%
%
\appendix

%
%
%
\subsection{Numerical Computation of the VLF Lower Bound}
\label{sec:bound_comp}

For channels with bounded information density, Wald's  equality (also known as Wald's identity or Wald's lemma) allows us to compute an upper bound on the expected stopping time $\E[\tau]$ in the random-coding lower bound of Thm.~\ref{thm:achievability} as follows:
\begin{align}
	\E[\tau] \leq \frac{\log (M-1) + \log \frac{1}{\epsilon} + B}{C}, 	\label{eqn:Wald_QED}
\end{align}
where $B < \infty$ is the upper bound on the information density:
%
$B = \sup \limits_{x \in {\cal X}, ~y \in \cal{Y}} i(x; y)$.
%
\begin{proof}
Defining $S_j = i(X_j; Y_j) = \log \frac{d\P(Y_j | X_j)}{d\P(Y_j)}$, we have
\begin{align}
	S^n = i(X^n; Y^n) &= \log \frac{d\P(Y^n | X^n)}{d\P(Y^n)} \\
	&= \sum \limits_{j=1}^n \log \frac{d\P(Y_j | X_j)}{d\P(Y_j)} 	\label{eqn:Wald_random_coding} \\
	&= \sum \limits_{j=1}^n S_j,
\end{align}
where \eqref{eqn:Wald_random_coding} follows due to the random codebook generation.
Since the $S_j$ are i.i.d with $\E[S_j] = C$, Wald's equality gives the following result \cite[Ch. 5]{Gallager_stochastic_textbook_2013} :
\begin{align}
	\E[S^\tau] &= \E[\tau]\E[S_1].
\end{align}
This leads to the following upper bound on $\E[\tau]$:
\begin{align}
	\E[\tau] &= \frac{\E[S^\tau]}{C} \\
	&= \frac{\E[S^{\tau-1} + S_\tau]}{C} \\
	&\leq \frac{\gamma + B}{C} 	\label{eqn:Wald_tau_def},
\end{align}
where \eqref{eqn:Wald_tau_def} follows from the definition of the threshold $\gamma$ in Thm.~\ref{thm:achievability} and of $B$ above.

Recall that the error probability $\epsilon$ is upper bounded in Thm.~\ref{thm:achievability} as $\epsilon \leq (M-1) \P[\bar \tau \leq \tau]$. This can be further upper bounded as follows, as in \cite{Polyanskiy_IT_2011_NonAsym} and \cite[Appendix B]{Williamson_ROVA_ISIT_2013}:
\begin{align}
	\P[\bar \tau \leq n] &= \E[\mathrm{1}\{\bar \tau \leq n\}] \\
	&= \E[\mathrm{1}\{\tau \leq n\} \exp\{-i(X^\tau; Y^\tau)]  \\
	&\leq \exp\{ -\gamma\},
\end{align}
because $i(X^\tau; Y^\tau) \geq \gamma$ by definition. Thus, we can write
\begin{align}
	\P[\bar \tau \leq \tau] &= \sum \limits_{n=0}^\infty \P[\tau = n] \P[\bar \tau \leq n] \\
	&\leq \sum \limits_{n=0}^\infty \P[\tau = n] \exp\{-\gamma\} \\
	&\leq \exp\{-\gamma\}.
\end{align}
Therefore, the bound on error probability can be loosened to $\epsilon \leq (M - 1) \exp\{-\gamma\}$.
Rearranging gives $\gamma \leq \log \frac{M-1}{\epsilon}$ and, when combined with \eqref{eqn:Wald_tau_def}, yields the following:
\begin{align}
	\E[\tau] &\leq \frac{\log \frac{M-1}{\epsilon} + B}{C},
\end{align}
which proves \eqref{eqn:Wald_QED}.
\end{proof}

\begin{proof}[\textbf{Examples}] For the BSC($p$), $B = \log 2(1 - p)$.
For the BI-AWGN channel, $B = \log 2$ (regardless of the SNR).
\end{proof}

%
%

\subsection{Proof of Cor.~\ref{thm:achievability_N} (Random-coding lower bound for repeat-after-$N$ codes)}
\label{sec:proof_achievability_N}

\begin{proof}
The proof closely follows that of \cite[Theorem 3]{Polyanskiy_IT_2011_NonAsym}. We define $M$ stopping times $\tau_j$, $j \in \{1,\ldots,M\}$, one for each codeword:
\begin{align}
	\tau_j = \inf \{n \geq 0:  i_N(X^n (j); Y^n) \geq \gamma \},
\end{align}
where $X^n (j)$ is the first $n$ symbols of the $j$th codeword. At each $n$, the decoder evaluates the $M$  information densities $i_N(X^n (j); Y^n)$ and makes a final decision at time $\tau^*$ when the first of these (possibly more than one at once) reaches the threshold $\gamma$:
\begin{align}
	\tau^* =\min \limits_{j=1, \dots, M} \tau_j .
\end{align}
The decoder at time $\tau^*$ selects codeword $m = \max \{j : \tau_j = \tau^* \}$.
The average blocklength $\ell = \mathrm{E}[\tau^*]$  is upper bounded as follows:
\begin{align}
	\mathrm{E}[\tau^*] &\leq \frac{1}{M} \sum \limits_{j=1}^M \mathrm{E}[\tau_j | W = j] \\
	&= \mathrm{E}[\tau_1 | W = 1] 		\label{eqn:vlf_N_proof_symm} \\
	&= \mathrm{E}[\tau] 			\label{eqn:vlf_N_proof_def} \\
	& = \sum \limits_{n=0}^\infty \P[\tau > n] \\
	& = \big(1 + \P[\tau > N] + \P[\tau > N]^2 + \dots \big) \sum \limits_{n=0}^{N-1} \P[\tau > n] 	\label{eqn:vlf_N_proof_repeat} \\
	& = \frac{\sum \nolimits_{n=0}^{N-1} \P[\tau > n]}{1 - \P[\tau > N]}. \label{eqn:vlf_N_proof_Etau}
\end{align}
Eq.~\eqref{eqn:vlf_N_proof_symm} follows from the symmetry of the $M$ stopping times and \eqref{eqn:vlf_N_proof_def} is by the definition of $\tau$ as given in \eqref{eqn:vlf_tau_N}.
Because the modified information densities depend only on the symbols in the current $N$-block, repeat-after-$N$ VLF codes satisfy the following property, which leads to \eqref{eqn:vlf_N_proof_repeat}:
\begin{align}
	\P[\tau > n] = \P[\tau > N]^s \P[\tau > r],
\end{align}
where $n = s N + r$.
The condition that $\P[\tau \leq N] > 0$ in Cor.~\ref{thm:achievability_N} is required so that $\P[\tau~>~N]~<~1$, guaranteeing that the sum in \eqref{eqn:vlf_N_proof_repeat} will converge.

Using $\hat W_n$ to denote the decoder's decision at time $n$, an error occurs if the decoder chooses $\hat W_{\tau^*} \neq W$. The probability of error $\epsilon$ can be bounded due to the random codebook generation:
\begin{align}
	\epsilon &= \P[\hat W_{\tau^*} \neq W] \\
	& \leq \P[\hat W_{\tau^*} \neq 1 | W = 1] \\
	& \leq \P[\tau_1 \geq \tau^*  | W = 1] \\
	& \leq \P \big[ \mathop{\cup}\limits_{j=2}^M \{ \tau_1 \geq \tau_j \} | W = 1 \big ] \\
	& \leq (M-1) \P [\tau_1 \geq \tau_2  | W = 1 ] \\
	& = (M-1) \P [\tau \geq \bar \tau ].
\end{align}
The last line follows because $(\tau, \bar \tau)$ have the same distribution as $(\tau_1, \tau_2)$ conditioned on $W = 1$.
\end{proof}

%
%
%
\subsection{General Blocklength-selection Algorithm}
\label{sec:blocklength_algo}


Selecting the $m$ incremental transmission lengths $\{I_i\}^*$ that minimize the latency (or equivalently, maximize the throughput) of VLF coding schemes is non-trivial. 
The complexity of a brute-force search grows exponentially with $m$.
In this section, we describe an efficient blocklength-selection algorithm that can be used to identify suitable blocklengths for general incremental redundancy schemes.
The goal of the algorithm is to select the $m$ integer-valued incremental transmission lengths $\{I_i\}^*$ as follows:
\begin{equation}
	\{I_i\}^* =\arg  \min \limits_{\{I_i\} \in \mathbb{Z} } \lambda ~ \text{ s.t. } ~P_\text{UE} \leq \epsilon.
	\label{eqn:objective}
\end{equation}
For the decision-feedback scheme using the TB ROVA in \secref{sec:stop_feedback}, the probability of undetected error is less than $\epsilon$ by definition, so the constraint can be ignored.

The proposed blocklength-selection algorithm for an $m$-transmission incremental redundancy scheme follows.
Starting from a pseudo-random initial vector $\{I_1, I_2, \dots, I_m\}$, the algorithm performs coordinate descent, wherein one transmission length $I_i$ is optimized while all others are held fixed. 
The objective function $\lambda$ is evaluated for positive and negative unit steps in increment $I_i$, i.e., for the transmission length vectors $\{I_1, \dots, I_i + 1 \dots, I_m\}$ and $\{I_1, \dots, I_i - 1 \dots, I_m\}$. Length $I_i$ is updated if the objective improves.
Once the objective cannot be improved by any single-coordinate steps, diagonal steps from each of the possible two-coordinate pairs $(I_i, I_j)$ are evaluated.
For each two-coordinate pair, four possible neighboring diagonal steps are evaluated. The transmission lengths $I_i$ and $I_j$ are updated if the best of the four diagonal steps improves the objective $\lambda$. This continues until the objective cannot be improved by additional diagonal steps.
The entire process then starts over from another pseudo-random initial vector. Random restarts are employed in order to avoid getting stuck at local optima of $\lambda$, of which there can be many. 
Empirical trials of this algorithm for several different families of retransmission probabilities demonstrated significantly reduced computation time compared to a brute-force approach (which is in general not possible for large $k$). Furthermore, while this algorithm is not guaranteed to find the global optimum, results show that the final objective value $\lambda_\text{best}$ was improved compared to results from an earlier quasi-brute-force trial.

%
\subsection{TB ROVA Version of Blocklength-selection Algorithm}
\label{sec:algo_tb_rova}

For the decision-feedback scheme of \secref{sec:stop_feedback} that uses the TB ROVA to compute the posterior probability of the decoded word, the probability of retransmission $\P_\text{re}(N)$ is the probability that the posterior at blocklength $N$ is less than $(1 - \epsilon)$, which is difficult to determine analytically.\footnote{For a given convolutional code, the weight spectrum for each blocklength $N$ could be used to bound or approximate the posterior probability and that could be used to bound or approximate the retransmission probability, but spectrum-based approaches tend not to be tight over a wide range of SNRs. Further complicating the task is the weight spectrum must be based on a rate-compatible puncturing pattern. Instead of optimizing this puncturing pattern, we use the same pseudo-random puncturing pattern throughout.}
Instead, we obtained estimates of the empirical retransmission-probabilities of the rate-$1/3$ convolutional codes in Table~\ref{tbl:vlf_conv_codes} and used those estimates of $\P_\text{re}(N)$ to compute the objective $\lambda$ in the algorithm.
To do so, we first simulated fixed-length transmission of rate-$1/3$, tail-biting convolutional codes from Table~\ref{tbl:vlf_conv_codes} at a small number of pseudo-randomly punctured blocklengths $N_\text{sim}$ for each fixed message-size $k$, where $N_\text{sim} \in \{k, k+\frac{k}{4}, k+\frac{2k}{4},\ldots, k+\frac{11k}{4}, 3k\}$.  For each $(k,N_\text{sim})$ pair, we counted the number of decoded words with posterior probability less than $(1 - \epsilon)$, indicating that a retransmission would be required, until there were at least 100 codewords that would trigger a retransmission. We computed $\P_\text{re}(N_\text{sim})$ according to 
\begin{align}
	\P_\text{re}(N_\text{sim}) = \frac{\text{\# codewords triggering retransmission}}{\text{total \# codewords simulated}}.
\end{align}
The full set of estimated retransmission probabilities $\tilde \P_\text{re}(N)$ for $N \in \{1, \dots, 3k\}$ was then obtained by a log-polynomial interpolation of the simulated values of $\P_\text{re}(N_\text{sim})$.
Finally, the estimated probabilities were used in the algorithm from \appref{sec:blocklength_algo} to select the optimal transmission lengths $\{I_i\}^*$.
We used 100 random restarts in our implementation. 
The performance of the TB ROVA-based retransmission scheme using these $m$$=$$5$ optimal blocklengths is evaluated in \secref{sec:stop_feedback}.

%
%
%

\subsection{Sampling Methodology}
\label{sec:sampling}

This section provides details about the number of codewords simulated and describes how the estimates of word-error probability, latency, and throughput were obtained. In particular, we show that in order to arrive at reliable estimates of the latency and throughput, tallying about 25 word errors is sufficient, in contrast to the well-known heuristic of tallying 100 word errors.

The Monte Carlo estimate $\hat P_{\text{UE}}$ of the undetected word-error probability $P_{\text{UE}}$ is 
\begin{align}
	\hat P_{\text{UE}} = \frac{1}{S} \sum \limits_{j=1}^S Z_j,	\label{eqn:P_UE_est}
\end{align}
where $S$ is the number of samples (i.e., the number of codewords simulated) and $Z_j$ is a Bernoulli indicator variable for the $j$th trial according to
\begin{align}
	Z_j =  \begin{cases} 1 &\mbox{error on trial } j \\
	0 & \mbox{success on trial } j . \end{cases} 
\end{align}
%
This Monte Carlo estimator is unbiased.
The variance of each Bernoulli random variable $Z_j$ is $\sigma_\text{UE}^2 := P_\text{UE} - P_\text{UE}^2$.
The sample variance of the estimator $\hat P_\text{UE}$ is $\hat \sigma_{\text{UE}}^2 := \hat P_\text{UE} - \hat P_\text{UE}^2$.

\begin{figure}[th]
  \centering
\ifx \columnmode \single
    \scalebox{0.6}{\includegraphics{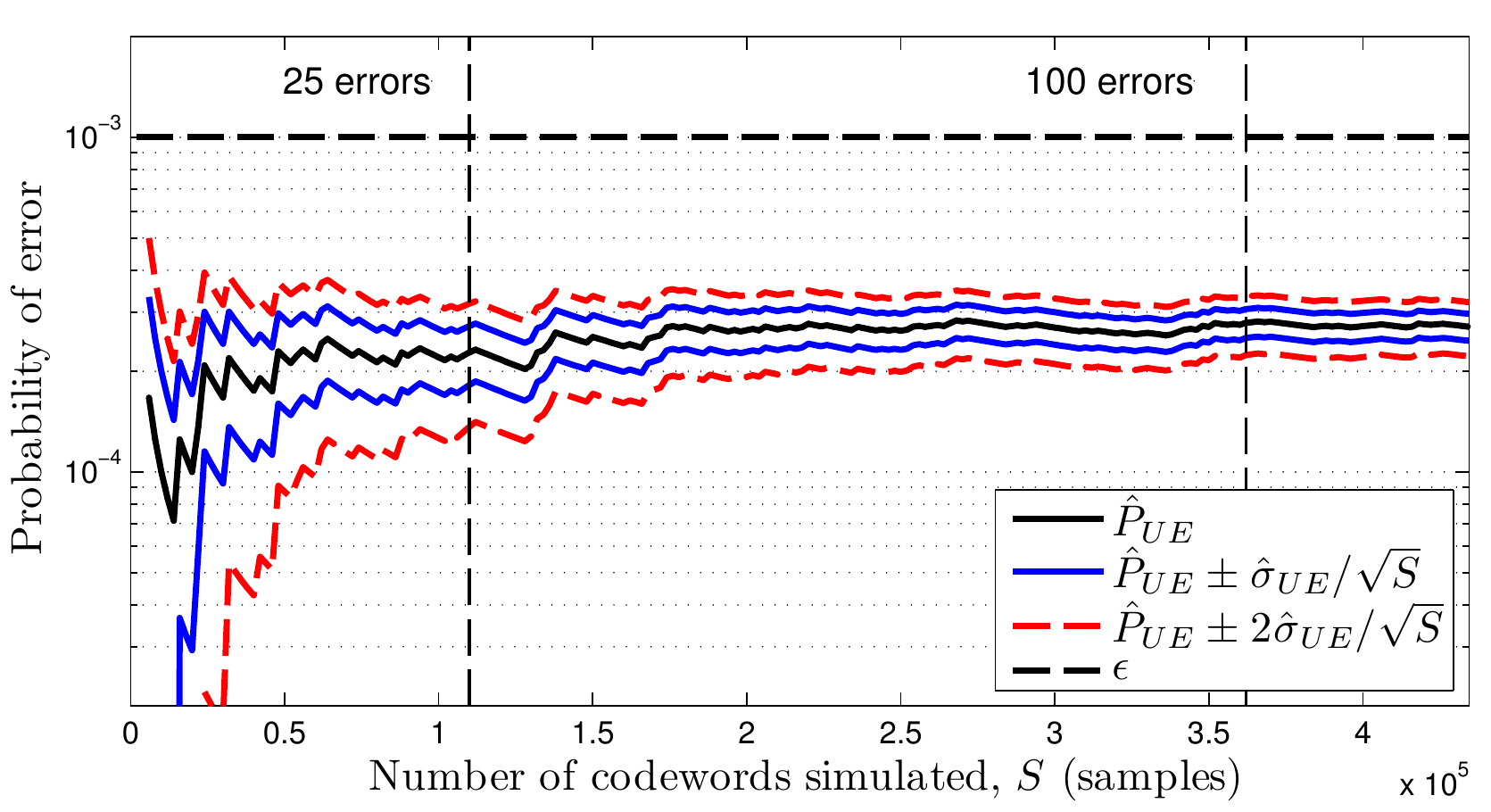}}
\else
    \scalebox{0.5}{\includegraphics{willi7.pdf}}
\fi
    \hspace{0.0in}
  \caption{One realization of the error probability estimator $\hat P_\text{UE}$ as a function of the number of independent trials $S$. Included in the figure are the estimates plus and minus one and two normalized sample standard deviations. The two dashed vertical lines correspond to the number of trials when 25 and 100 word-errors were accumulated.}
   \label{fig:sampling_WEP}
\end{figure}

\begin{figure}
  \centering
\ifx \columnmode \single
    \scalebox{0.6}{\includegraphics{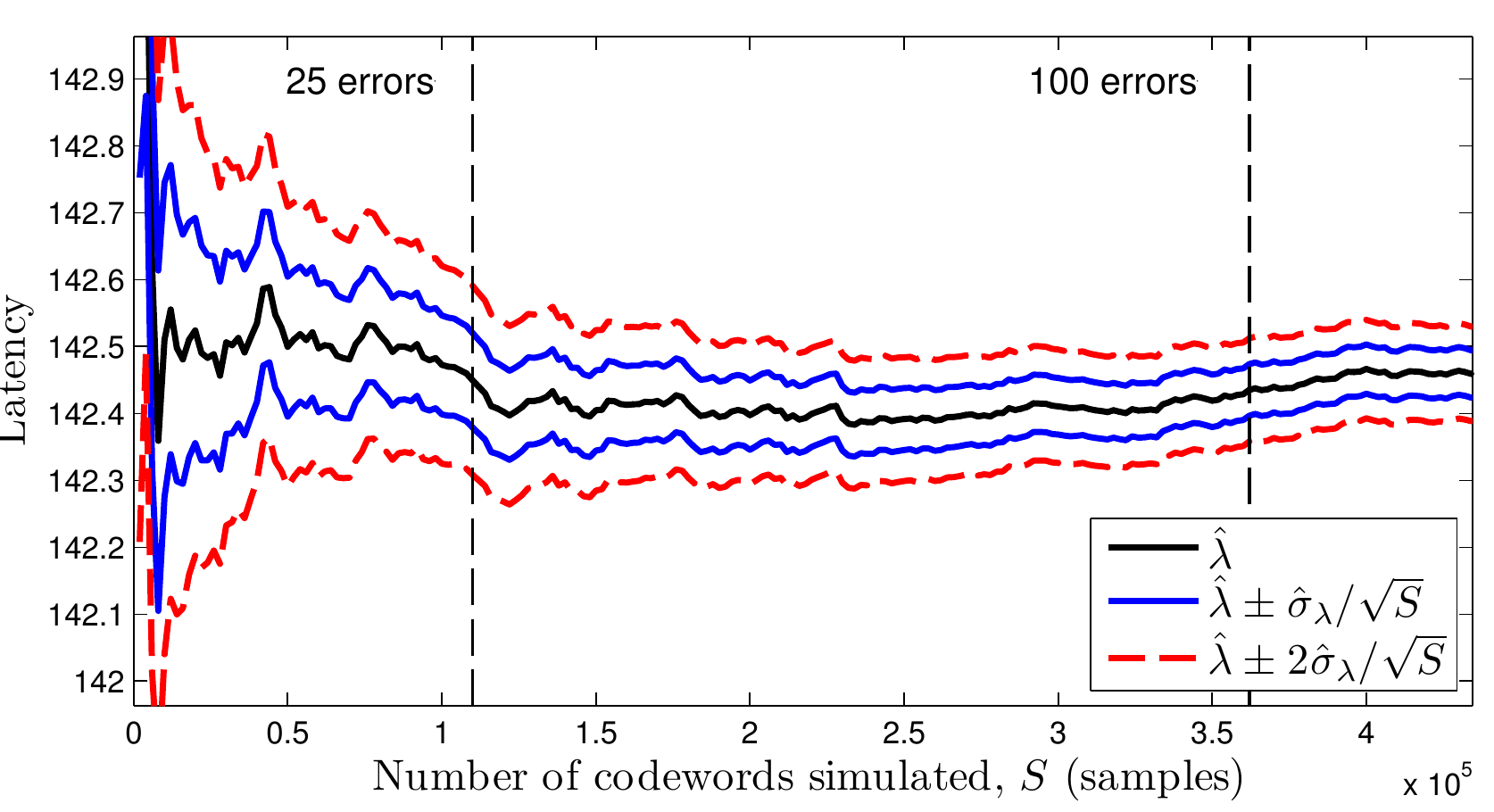}}
\else
    \scalebox{0.45}{\includegraphics{willi8.pdf}}
\fi
    \hspace{0.0in}
  \caption{One realization of the latency estimator $\hat \lambda$ as a function of the number of independent trials $S$. Included in the figure are the estimates plus and minus one and two normalized sample standard deviations. The two dashed vertical lines correspond to the number of trials when 25 and 100 word-errors were accumulated.}
   \label{fig:sampling_latency}
\end{figure}

Fig.~\ref{fig:sampling_WEP} illustrates one realization of the word-error probability estimate $\hat P_\text{UE}$ and the normalized sample standard deviation $\hat \sigma_\text{UE} / \sqrt{S}$. Both Figs.~\ref{fig:sampling_WEP} and \ref{fig:sampling_latency} correspond to the 64-state TBCC with $k$$=$$64$ information bits with $m$$=$$5$ transmissions, for $\epsilon$$=$$10^{-3}$ and AWGN SNR 2 dB.

A common heuristic is to run simulations until there are $100$ word errors, or until $S \approx 100/P_\text{UE} $. This can be explained in terms of confidence intervals as follows, as described in Dolecek et al. \cite{Dolecek_Importance_Sampling_ITW_2007}. Suppose we want the error of the estimator to be less than fraction  $\beta$ of $P_\text{UE}$ with confidence level $c$. That is, we want
\begin{align}
	\P \left[| \hat P_\text{UE} - P_\text{UE} | > \beta P_\text{UE} \right] &\leq 1 - c.	\label{eqn:estimator_error}
\end{align}
Appealing to the Central Limit Theorem as the number of samples $S$ becomes large, because $\hat P_\text{UE}$ is a sum of independent Bernoulli random variables, as in \eqref{eqn:P_UE_est}, $\hat P_\text{UE}$ converges to a Gaussian random variable with mean $P_\text{UE}$ and variance $\sigma_\text{UE}^2 / S $. Thus, we can approximate the left-hand side of \eqref{eqn:estimator_error} as the tail probability of a standard normal random variable:
\ifx \columnmode \single
\begin{align}
	\P [| \hat P_\text{UE} - P_\text{UE} | > \beta P_\text{UE} ] &= \P \left[ \frac{| \hat P_\text{UE} - P_\text{UE} |}{\sqrt{P_\text{UE} (1 - P_\text{UE})/S}} > \beta \sqrt{\frac{S P_\text{UE}}{1 - P_\text{UE}}} \right ] \\
	&\approx \P \left[ | Y | > y  \right],
\end{align}
\else
\begin{align}
	\P [ &| \hat P_\text{UE} - P_\text{UE} | > \beta P_\text{UE} ] \nonumber \\ 
	&= \P \left[ \frac{| \hat P_\text{UE} - P_\text{UE} |}{\sqrt{P_\text{UE} (1 - P_\text{UE})/S}} > \beta \sqrt{\frac{S P_\text{UE}}{1 - P_\text{UE}}} \right ] \\
	&\approx \P \left[ | Y | > y  \right],
\end{align}
\fi
where $Y \sim {\cal N}(0,1)$ and $y = \beta \sqrt{\frac{S P_\text{UE}}{1 - P_\text{UE}}}$.
In order to, for example, have confidence level $c = 95\%$ (corresponding to $y\approx2$) and $\beta = 0.2$ (corresponding to a confidence interval $[0.8 P_\text{UE}, 1.2 P_\text{UE}]$), we need approximately $S = 100 (1 - P_\text{UE})/P_\text{UE} \approx 100/P_\text{UE}$ samples.

However, the construction of the reliability-based stopping rule in \eqref{eqn:rova_tau} guarantees that the word-error probability is less than $\epsilon$, which satisfies the error requirement for an $(\ell,M,\epsilon)$ VLF code. 
With that guarantee in mind, the goal of simulations in this paper is not to characterize the error probability, but rather to estimate the latency and throughput of VLF codes at short blocklengths. It is therefore more instructive to analyze the sample variance for the latency, as shown below.
%
%

The Monte Carlo estimate $\hat \lambda$ of the latency $\lambda$ is
\begin{align}
	\hat \lambda = \frac{1}{S} \sum \limits_{j=1}^S \tau_j,
\end{align}
where $\tau_j \in \{N_1, N_2, \dots, N_m, N_m + N_1, \dots \}$ is the cumulative number of transmitted symbols (i.e., the blocklength) in the $j$th trial.
Noting that the expected latency is $\lambda = \E[\tau]$, the variance of the latency is $\sigma_\lambda^2 := \E[\tau^2 - \lambda^2]$. Neither $\lambda$ nor $\sigma_\lambda^2$ is known, but $\hat \lambda$ is an unbiased estimator for $\lambda$ with little variance.
The sample variance of the estimator $\hat \lambda$ is 
\begin{align}
	\hat \sigma_{\lambda}^2 &:= \left( \frac{1}{S} \sum \limits_{j=1}^S \tau_j^2 \right) - \hat \lambda^2.
\end{align}
The throughput $R_t$ is estimated as $\hat R_t = k (1 - \hat P_\text{UE}) / \hat \lambda$.

Fig.~\ref{fig:sampling_latency} provides an example of one realization of the latency estimate $\hat \lambda$ versus the number of independent trials $S$.
The normalized sample standard deviation $\hat \sigma_\lambda / \sqrt{S}$ is also included.
As can be seen from Fig.~\ref{fig:sampling_latency}, once $S\approx 25 / {\hat P_\text{UE}}$ samples have been drawn (see the dashed vertical line labeled `25 errors'), the standard deviation of the latency estimate is quite small relative to the estimate itself (less than $1\%$), indicating that estimates of both the latency and throughput given in Sec.~\ref{sec:stop_feedback} are sufficiently reliable.  This is not surprising since with $\epsilon$$=$$10^{-3}$ we are using at least 25,000 independent observations of the random variable $\tau$ to compute its mean.
%

%
%
\bibliographystyle{IEEEtran}
{\bibliography{AW_bib}}

\begin{IEEEbiographynophoto}{Adam Williamson SM'12-M'14}
Adam Williamson received the B.S. in Electrical and Computer Engineering, B.S. in Applied Math, and B.A. in Physics from the University of Rochester, all in 2008. He received the M.S. and Ph.D. in Electrical Engineering from the University of California, Los Angeles in 2012 and 2014, respectively, where he was part of the Communication Systems Laboratory.
Adam's research interests include feedback coding, hybrid ARQ and energy-efficient communication. He is now with Northrop Grumman.
\end{IEEEbiographynophoto}

\begin{IEEEbiographynophoto}{Tsung-Yi Chen SM'11-M'13}
Tsung-Yi Chen received his B.S. degree in Electrical Engineering from National Tsing Hua University, Taiwan, in 2007. He obtained his M.S. and Ph.D. degrees in Electrical Engineering from UCLA in 2009 and 2013, respectively. He is a recipient of the UCLA Dissertation Year Fellowship 2012-2013. He joined Northwestern University, Evanston in 2013 as a postdoctoral fellow.
His research includes coding theory and information theory, with applications to feedback communication, flash memory storage systems, and machine-to-machine communication. 
\end{IEEEbiographynophoto}

\begin{IEEEbiographynophoto}{Richard Wesel SM'91-M'96-SM'01 }
Richard D. Wesel is a Professor with the UCLA Electrical Engineering Department and
is the Associate Dean for Academic and Student Affairs for the UCLA Henry Samueli 
School of Engineering and Applied Science. He joined UCLA in 1996 after receiving 
his Ph.D. in electrical engineering from Stanford. His B.S. and M.S. degrees in electrical 
engineering are from MIT. His research is in the area of communication theory with 
particular interest in channel coding. He has received the National Science Foundation 
(NSF) CAREER Award, an Okawa Foundation award for research in information theory 
and telecommunications, and the Excellence in Teaching Award from the Henry Samueli 
School of Engineering and Applied Science.
\end{IEEEbiographynophoto}

\newpage

\newpage

\end{document}